\documentclass[aip,cha,twocolumn,reprint]{revtex4-1}
\usepackage{graphicx,mathtools,amsmath,subfigure,amssymb,comment,tabularx}

\usepackage[usenames,dvipsnames,svgnames,table]{xcolor}
\definecolor{black}{rgb}{0,0,0}
\definecolor{myblue}{rgb}{0,0,0.6}
\definecolor{myred}{rgb}{0.6,0,0}

\newcommand{\dd}{\mathrm{d}}

\newcommand{\APD}{\mathrm{APD}}
\newcommand{\DI}{\mathrm{DI}}

\renewcommand{\vec}[1]{\mathbf{#1}}

\begin{document}

\title{Dynamical mechanism of atrial fibrillation: a topological approach}

\author{Christopher D. Marcotte}
\affiliation{EPSRC Centre for Predictive Modelling in Healthcare, University of Exeter, EX44QJ, UK}
\author{Roman O. Grigoriev}
\affiliation{School of Physics, Georgia Institute of Technology, Atlanta, Georgia, 30332-0430, USA}

\date{\today}

\begin{abstract}
While spiral wave breakup has been implicated in the emergence of atrial fibrillation, its role in maintaining this complex type of cardiac arrhythmia is less clear.
We used the Karma model of cardiac excitation to investigate the dynamical mechanisms that sustain atrial fibrillation once it has been established.
The results of our numerical study show that spatiotemporally chaotic dynamics in this regime can be described as a dynamical equilibrium between topologically distinct types of transitions that increase or decrease the number of wavelets, in general agreement with the multiple wavelets hypothesis.
Surprisingly, we found that the process of continuous excitation waves breaking up into discontinuous pieces plays no role whatsoever in maintaining spatiotemporal complexity. 
Instead this complexity is maintained as a dynamical balance between wave coalescence -- a unique, previously unidentified, topological process that increases the number of wavelets -- and wave collapse -- a different topological process that decreases their number.
\end{abstract}


\maketitle

\begin{quotation}
Atrial fibrillation is a type of cardiac arrhythmia featuring multiple wavelets that continually interact with each other, appear, and disappear.
The genesis of this spatiotemporally chaotic state has been linked to the alternans instability that leads to conduction block and wave breakup, generating an increasing number of wavelets.
Less clear are the dynamical mechanisms that sustain this state and, in particular, maintain the balance between the creation and destruction of spiral wavelets.
Even the relation between wave breakup and conduction block, which is well-understood qualitatively, at present lacks proper quantitative description.
This paper introduces a topological description of spiral wave chaos in terms of the dynamics of wavefronts, wavebacks, and point defects -- phase singularities -- that anchor the wavelets.
This description both allows a dramatic simplification of the spatiotemporally chaotic dynamics and enables quantitative prediction of the key properties of excitation patterns.
\end{quotation}

\section{Introduction}

Atrial fibrillation (AF) is the most common sustained cardiac arrhythmia \cite{Waktare1998}. While not itself lethal, it has a number of serious side effects, such as increased risk of stroke and systemic thromboembolism \cite{Noel1991}.
The origin of AF has been debated through much of the previous century \cite{Nattel2000}. 
In 1913, Mines proposed that fibrillation is caused by a reentrant process \cite{mines1913}, which leads to a high-frequency wave propagating away from the reentry site and breaking up into smaller fragments.
This mechanism is presently referred to as anatomical reentry and requires a structural heterogeneity of the cardiac tissue, such as a blood vessel (e.g., vena cava).
Reentry could also be functional \cite{Allessie1977}, where the heterogeneity (i.e., tissue refractoriness) is dynamical in nature.
Neither picture, however, explains the complexity and irregularity of the resulting dynamics. 
 
The first qualitative explanation of this complexity came in the form of the multiple wavelet hypothesis proposed by Moe \cite{Moe1962}.
In this hypothesis multiple independent wavelets circulate around functionally refractory tissue, with some wavelets running into regions of reduced excitability and disappearing and others breaking up into several daughter wavelets, leading to a dynamical equilibrium.
This picture was subsequently evaluated and refined based on numerical simulations \cite{Moe1964} and experiments \cite{Allessie1985}. 

Krinsky \cite{krinskii1966spread} and then Winfree \cite{Winfree1987} suggested that the dynamical mechanism of fibrillation relies on the formation and interaction of spiral waves.
Spiral waves rotate around phase singularities that may or may not move, producing reentry that requires neither structural nor dynamical heterogeneity in refractoriness.
The presence and crucial role of spiral waves in AF was later confirmed in optical phase mapping experiments \cite{davidenko1992,Pertsov1993,Chen2000}.
Experimental evidence shows that spiral waves tend to be very unstable: only a small fraction of these complete a full rotation \cite{Chen1998comp,Lee2001,Liu2003} with the dynamics dominated by what appears to be wavebreaks or wave breakups (WBs). 

Although there is plentiful experimental and computational \cite{Fenton2002} evidence that WBs play a crucial role in the {\it transition} to fibrillation, it is far less obvious that this mechanism is essential for {\it maintaining} AF.
As Liu {\it et al.} \cite{Liu2003} write, ``for a wave to break, its wavelength must become zero at a discrete point somewhere along the wave. 
This can happen if the wave encounters refractoriness that creates local block (wavelength = 0), while propagating elsewhere. 
Therefore, WBs can be detected at locations where activating wavefronts meet the repolarization wavebacks.''
The data produced by experimental studies is highly unreliable in this regard, since detecting the position of wavefronts and wavebacks based on optical recordings is far from straightforward.
Numerical simulations, on the other hand, have focused mostly on the transition, rather than sustained AF.
Theoretical studies of model systems such as the complex Ginzburg-Landau equation \cite{Aranson2002} and FitzHugh-Nagumo equation \cite{Panfilov1993} lack dynamical features, such as the alternans instability \cite{Nolasco1968}, that are believed to play an essential role in conduction block that leads to WBs and AF \cite{Narayan2011}.

Even if WBs do play a role in maintaining AF, they tell only a part of the story. 
Indeed, in sustained AF, despite some variation, the quantitative metrics such as the number of wavelets or phase singularities, have to remain in dynamical equilibrium.
While WBs may explain the increase in the number of wavelets and phase singularities, it cannot explain how these numbers might ever decrease.
The multiple wavelet hypothesis \cite{Moe1962} comes the closest to providing all the necessary ingredients for such a dynamical equilibrium, but it lacks sufficient detail to be either validated or refuted.

The main objective of this paper is to construct a mathematically rigorous topological description of the dynamics and to use this description to characterize and classify different dynamical events that change the topological structure of the pattern of excitation waves in a state of sustained atrial fibrillation.
We will focus on the smoothed version \cite{ByMaGr14} of the Karma model~\cite{Karma1993,karma94},
\begin{equation}\label{eq:rde}
	\partial_t\vec{u} = D\nabla^{2}\vec{u} + \vec{f}(\vec{u}),
\end{equation}
where $\vec{u}(t,\vec{x}) = [u_1,u_2](t,\vec{x})$,
\begin{align}\label{eq:karmakinetics}
	f_1&=(u^* - u_2^{M})\{1 - \tanh(u_1-3)\}u_1^{2}/2 - u_1,\\
	f_2&=\epsilon\left\{\beta \Theta_s(u_1-1) + \Theta_s(u_2-1)(u_2-1) - u_2\right\},\nonumber
\end{align}
where $\Theta_s(u)=[1+\tanh(su)]/2$, $u_1$ is the (fast) voltage variable, and $u_2$ is the (slow) gating variable. 
The parameter $\epsilon$ describes the ratio of the corresponding time scales, $s$ is the smoothing parameter, and the diagonal matrix $D$ of diffusion coefficients describes spatial coupling between neighboring cardiac cells (cardiomyocytes). 
{The parameters of the model~\cite{Marcotte2016b} are $M=4$, $\epsilon = 0.01$, $s = 1.2571$, $\beta = 1.389$, $u_1^{*} = 1.5415$, $D_{11} = 4.0062$, and $D_{22} = 0.20031$.}
This is the simplest model of cardiac tissue that develops sustained spiral wave chaos from an isolated spiral wave through the amplification of the alternans instability resulting in conduction block and wave breaks, mirroring the transition from tachycardia to fibrillation.

The outline of the paper is as follows:
Section \ref{sec:anatomy} introduces the topological description of the complicated multi-spiral states.
Section \ref{sec:block} discusses the relationship between tissue refractoriness and conduction block.
The dynamical mechanisms that maintain spiral wave chaos are presented in Section \ref{sec:results}. 
Section \ref{sec:spacing} discusses the statistical measures quantifying sustained dynamics, 
and Section \ref{sec:summary} contains the discussion of our results and conclusions.

\section{Wave anatomy}\label{sec:anatomy}

To quantitatively describe the topological changes such as wave breakup and creation/destruction of spiral {cores} we must first discuss the anatomy of excitation waves and define the appropriate terminology.

\subsection{Wavefront and waveback}

The region of excitation can be thought of as being bounded by the wavefront, which describes fast depolarization of cardiac cells, and waveback, which describes a typically slower repolarization. The most conventional definition of the wavefront and waveback that has been used in both experiment \cite{Lee2001,roberts1979influence} and numerical simulations \cite{Fenton2002,murthy2013curvature} is based on a level-set of the voltage variable, $u_1(t,\vec{x}) = \bar{u}_1$.
If we define the region of repolarization
\begin{equation}\label{eq:R}
R=\{ \vec{x} | \partial_tu_1(t,\vec{x})) < 0 \},
\end{equation}
then the wavefront/back is the part of the level set outside/inside $R$.

The choice of the voltage threshold $\bar{u}_{1}$ is arbitrary and is typically taken as a percentage of the difference between the voltage maximum and its value in the rest state~\cite{Fenton2002}.
This very simple definition allows easy identification of the action potential duration (APD) and diastolic interval ($\DI$), where the percentage is often used as a subscript which refers to the choice of the threshold (e.g., $\APD_{80}$ corresponds to 80\% of the difference).

\begin{figure}[tpb]
	\subfigure[]{\includegraphics[width=0.55\columnwidth]{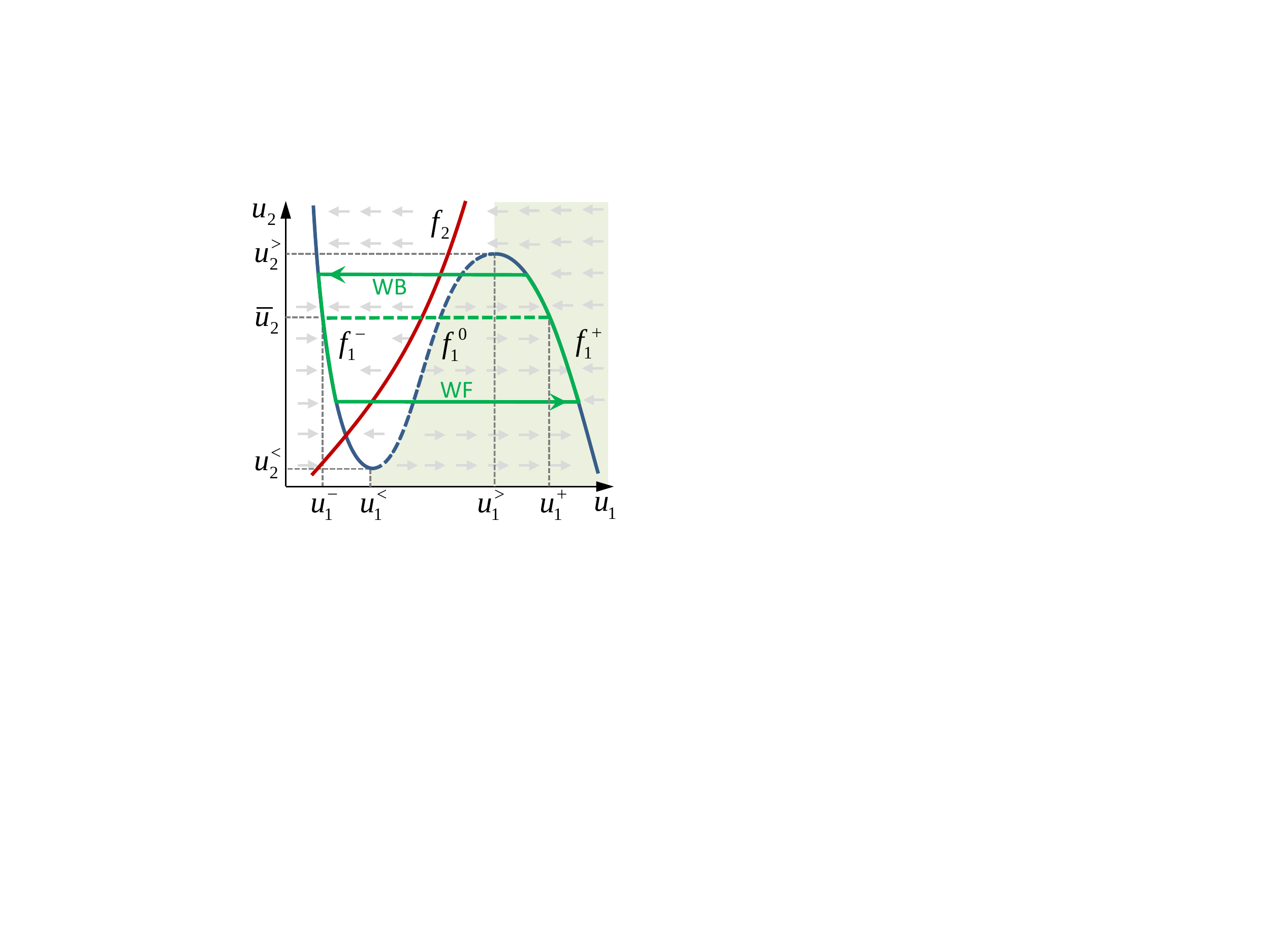}}
	\subfigure[]{\includegraphics[width=0.43\columnwidth]{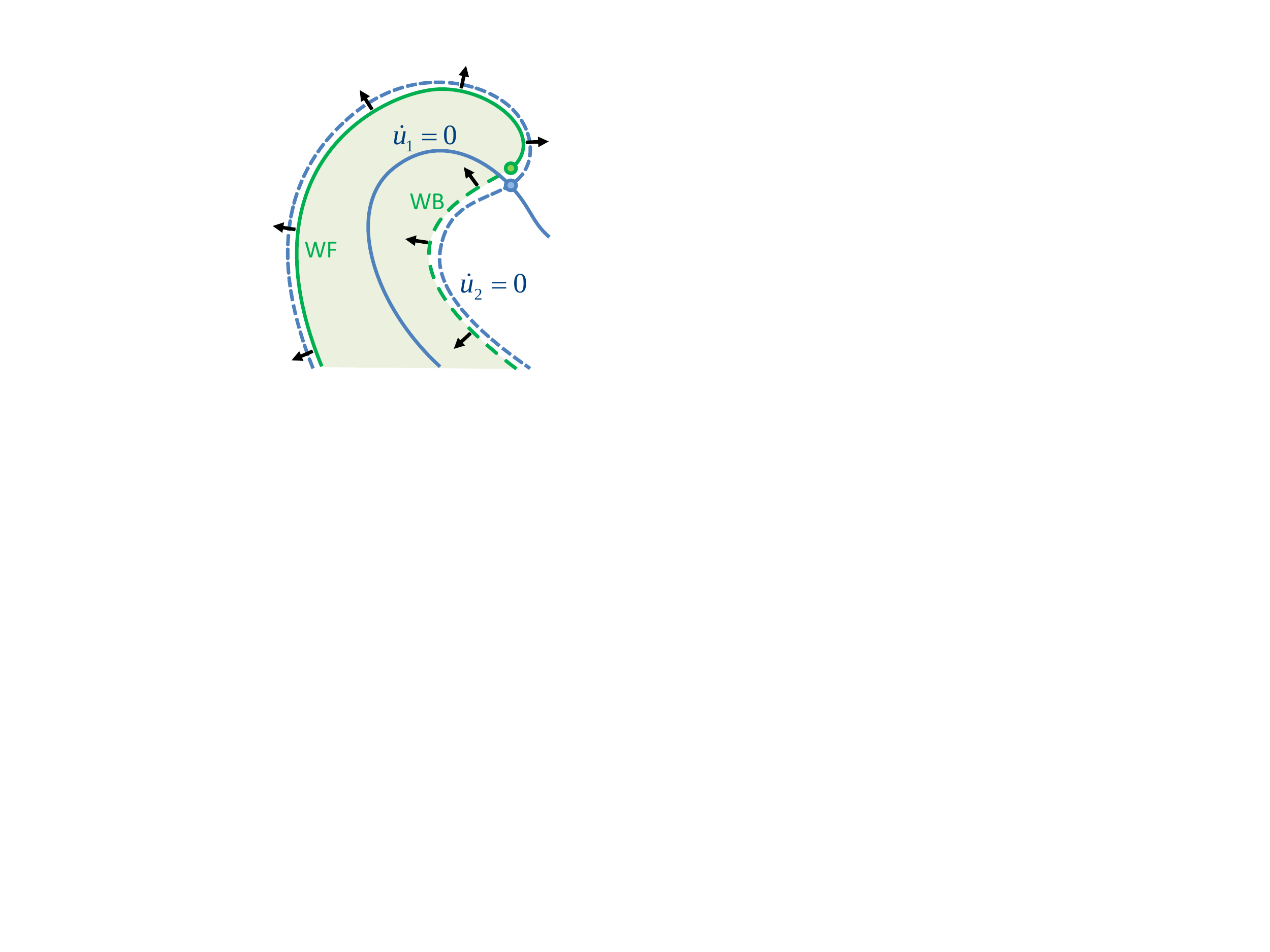}}
	\caption{Phase diagram for a generic two-variable model in the $\epsilon\to0$, $D\to0$ limit. (a) The nullclines $f_1(\vec{u})=0$ (blue), $f_2(\vec{u})=0$ (red) and the limit cycle oscillation (solid green). 
	The labels WF and WB denote the wavefront and waveback, respectively. 
	The dashed green line corresponds to the ``stall'' value $\bar{u}_2$ which defines a stationary front. 
	The excited region $E$ is shaded, and the gray arrows show the local direction of the vector field ${\bf f}({\bf u})$. 
	(b) The wavefront (solid green line) and waveback (dashed green line), the level sets $\dot{u}_1=0$ (solid blue line) and $\dot{u}_2=0$ (dashed blue line), spiral wave tip (green circle) and phase singularity (blue circle). The arrows denote the direction of the normal velocity $c$.
	\label{fig:anatomy}}
\end{figure}

Analytical studies tend to use a different definition \cite{Tyson1980,Tyson1988,cross93} for the wavefront and waveback which is based on scale separation between the dynamics of fast variables, such as voltage, and slow variables, such as potassium concentration (the voltage $u_1$ and the gating variable $u_2$, respectively, in the Karma model). For the simplest two-variable models (Karma, Barkley~\cite{barkley1991model}, FitzHugh-Nagumo~\cite{FitzH61}, Rinzel-Keller~\cite{Rinzel1973}, etc.), {in the limit $\epsilon\to0$, the excitation waves are related to a limit cycle oscillation in the $(u_1,u_2)$ plane governed by the system of coupled ordinary differential equations \cite{cross93}}
\begin{align}\label{eq:ode}
	\dot{u}_1 &= f_1(\vec{u}),\nonumber\\
	\dot{u}_2 &= f_2(\vec{u}).
\end{align}
The wavefront and waveback correspond to the segments of the limit cycle solution of \eqref{eq:ode} connecting the two stable branches of the $u_1$-nullcline $f_1({\bf u})=0$ for which $\dd u_2/\dd u_1<0$ (they are denoted with superscripts $-$ and $+$, as illustrated in Fig. \ref{fig:anatomy}(a)). 

In the limit $\epsilon\to 0$ these segments become horizontal lines and describe very fast (in time) variation of the voltage variable. In space, both the wavefront and the waveback have widths that scale as {$\sqrt{\epsilon}$}. In the limit $\epsilon\to0$ they become very sharp (green curves in Fig. \ref{fig:anatomy}(b)) and can be thought of as the boundaries of the region of excited tissue 
\begin{equation}\label{eq:etbf}
E=\{ \vec{x} | f_1^+({\bf u}(t,\vec{x})) = O(\epsilon) \},
\end{equation}
shown as the green shaded area in Fig. \ref{fig:anatomy}(b). 
These boundaries can be defined with equal precision using any curve in the $(u_1,u_2)$ plane that bisects both the wavefront and the waveback segments of the limit cycle. If we define this curve as the zero level set of an indicator function
\begin{equation}\label{eq:lsu}
	g(\vec{u})=0,
\end{equation}
the wavefront and the waveback in the physical space at a particular time $t$ are given by \begin{equation}\label{eq:lsx}
\partial E=\{ \vec{x} | g(\vec{u}(t,\vec{x})) = 0 \}.
\end{equation}
In particular, a level set of the voltage variable discussed previously corresponds to a vertical line in Fig. \ref{fig:anatomy}(a), $g(\vec{u})=u_1-\bar{u}_1$. 

A less arbitrary and dynamically better justified choice $g(\vec{u})=f_1^0({\bf u})$ corresponds to the unstable branch of the $u_1$-nullcline (for which $\dd u_2 / \dd u_1 > 0$).
In order to generalize this choice to finite values of $\epsilon$, the unstable branch has to be extended beyond its end points ${\bf u}^\lessgtr$ where $\dd u_2 / \dd u_1 = 0$, e.g.,
\begin{equation}\label{eq:gf}
	g(\vec{u})=\left\{
	\begin{matrix}
	u_1-u_1^<,\ & u_2\leq u_2^<,\\
	f_1^0({\bf u}), & u_2^<<u_2<u_2^>,\\
	u_1-u_1^>,\ & u_2\geq u_2^>.\\
	\end{matrix}
	\right.
\end{equation}
The excited region where $g({\bf u})>0$ according to \eqref{eq:gf} is shaded green in Fig. \ref{fig:anatomy}(a). 
Yet another alternative suggested by Fig. \ref{fig:anatomy}(a) is to use the $u_2$-nullcline $g(\vec{u})=f_2({\bf u})$ that also bisects both the wavefront and the waveback.

As time evolves, the level set \eqref{eq:lsx} moves with normal velocity
\begin{equation}\label{eq:cn}
c=-\frac{\vec{b}\cdot\partial_t\vec{u}}{\vec{b}\cdot\partial_n\vec{u}},
\end{equation}
where $\vec{b}=\partial g/\partial\vec{u}$, $\partial_n=\vec{n}\cdot\nabla$ is the directional derivative, and $\vec{n}$ is the outside normal to $\partial E$. 
This normal velocity is taken to be positive for the wavefront and negative for the waveback.
In particular, for $g(\vec{u})=u_1-\bar{u}_1$, \eqref{eq:cn} simplifies, yielding $c=-\partial_tu_1/\partial_nu_1$, so the sign of $c$ corresponds to the sign of $\partial_tu_1$ {for proper choices of $\bar{u}_1$ such that $\partial_nu_1<0$ over the entire level set}.

Whatever the choice of the bisecting curve $g({\bf u})=0$ in the $(u_1,u_2)$ plane is, it may not define a continuous curve in the physical space at all times.
Indeed, the PDE model \eqref{eq:rde} does not take the cellular structure of the tissue into account.
Instead, a spatial discretization of \eqref{eq:rde} should be used, so that the field $\vec{u}$ becomes a discontinuous function of space. 
In this case the wavefront and waveback should instead be defined as the boundary $\partial E$ of the region
\begin{equation}\label{eq:etbg}
E=\{ \vec{x} | g(\vec{u}(t,\vec{x})) > 0 \},
\end{equation}
rather than the level set \eqref{eq:lsx}.

There are several serious problems with the ``local'' definitions discussed above, which are based on the kinetics of isolated cardiac cells. 
For one, they ignore the coupling between neighboring cells in tissue (electrotonic effects) and hence cannot correctly describe the essential properties of excitability and refractoriness, making quantitative description of conduction block impossible. 
Furthermore, since the value of {$\epsilon$} is not vanishingly small for physiologically relevant models, the widths of both the wavefront and the waveback become finite as well, so different choices of $g(\vec{u})$ can produce rather distinct results in the physical space. Additional complications will be discussed below.

\begin{figure}[htpb]
	\subfigure[]{\includegraphics[scale=0.6]{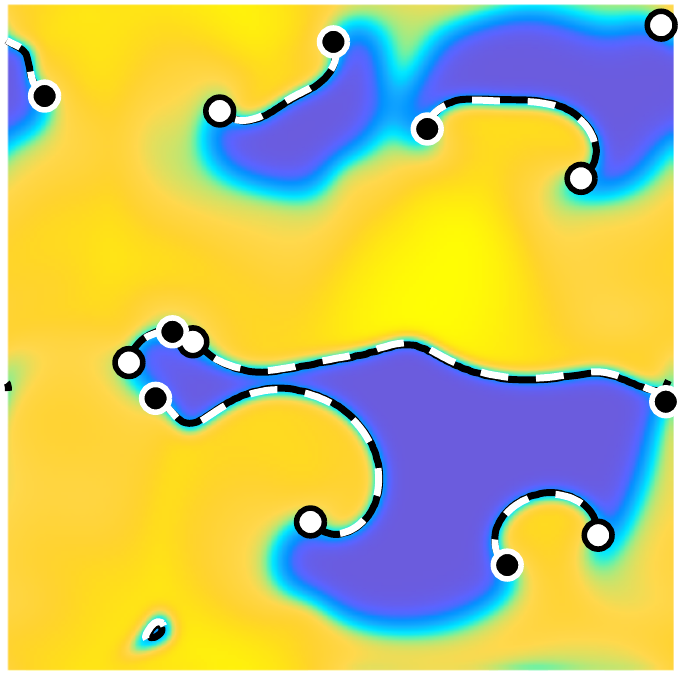}}
	\subfigure[]{\includegraphics[scale=0.6]{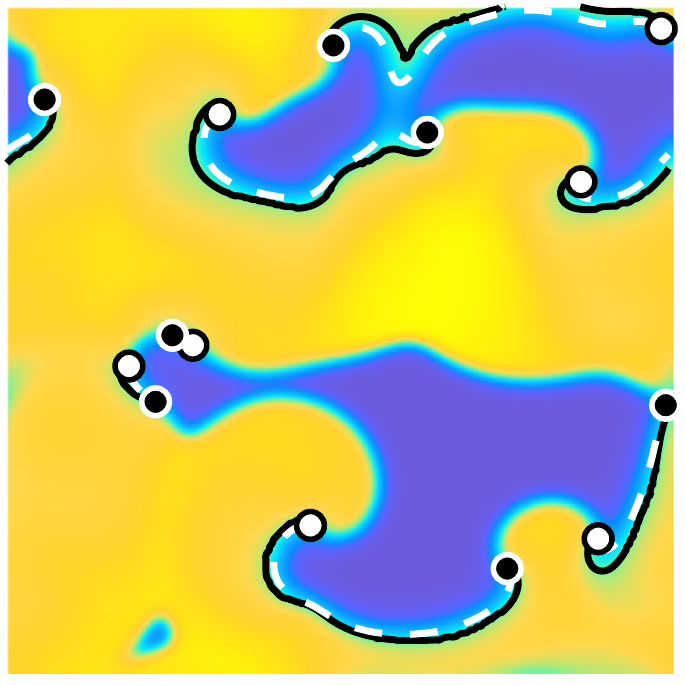}}
	\subfigure[]{\includegraphics[width=.7\columnwidth]{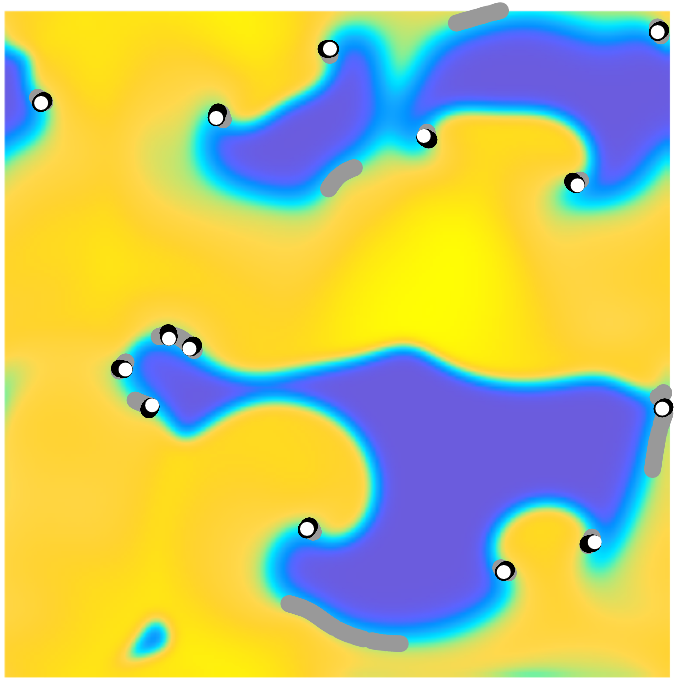}}
	\includegraphics[scale=0.85]{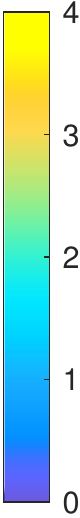}
	\caption{Comparison of the definitions of the wavefronts (a) and wavebacks (b) for a complicated multi-spiral state $u_{1}(t,\vec{x})$ based on \eqref{eq:lsx}-\eqref{eq:gf} (solid black lines) and \eqref{eq:dE} (dashed white lines). White/black circles correspond to the phase singularities with positive/negative chirality.
	(c) Comparison between {spiral wave tips and phase singularities} computed using definitions \eqref{eq:lsi}, \eqref{eq:znv}, and \eqref{eq:origins} shown as gray, black, and white points, respectively.
	\label{fig:WFWBcores}}
\end{figure}

The definitions of the wavefront and waveback can be generalized for a tissue model by noticing that the level sets such as $f_1^0(\vec{u})=0$ or $f_2(\vec{u})=0$, in the limit {$\epsilon\to 0$,} coincide with the level sets $\partial_tu_1=0$ and $\partial_tu_2=0$. These are special cases of a more general relation
\begin{equation}\label{eq:adu}
	g(\vec{u})=\vec{a}\cdot\partial_t\vec{u}=0,
\end{equation}
where $\vec{a}=(\cos\alpha,\sin\alpha)$ and {$0<\alpha<\pi$} is a parameter that can be chosen to {properly} describe the refractoriness and excitability {of the model} for finite values of $\epsilon$. 
The wavefront and waveback can again be distinguished as the parts of the level set that lie outside or inside $R$, respectively.
We will set $\alpha=\pi/2$ below, which yields the following definition
\begin{equation}\label{eq:dE}
	\partial E = \{ \vec{x} | \partial_t u_2(t,\vec{x}) = 0 \}.
\end{equation}
The level set \eqref{eq:dE} is shown as the dashed blue line in Fig. \ref{fig:anatomy}(b).
As Fig. \ref{fig:WFWBcores}(a-b) illustrates, for the Karma model, it gives an extremely good agreement with the more conventional definition based on \eqref{eq:gf} for both the wavefront and the waveback.
{Variation in $\alpha$ by $O(\epsilon)$ has} a very weak effect on the position of the wavefront (which is very {sharp}), but {has a more} pronounced effect on the position of the waveback (which is much broader).

\subsection{Phase singularities}

Typically (although certainly not always \cite{biktashev2005causodynamics}), the temporal frequency and wavelength of spiral waves are controlled by their central region, usually referred to as a spiral core or rotor~\cite{Winfree1978}. 
This region is spatially extended and its size can be characterized using the adjoint eigenfunctions of the linearization \cite{biktasheva1998,Marcotte2016}. 
In practice it is more convenient to deal with a single point that describes the location of the core region.
In particular, the center of this region is associated with a phase singularity, where the amplitude of oscillation vanishes.
{The location of the phase singularity depends on the definition of the phase, however, and the proper definition is far from obvious for strongly nonlinear oscillations characteristic of excitable systems.
The methods based on phase \cite{Gray1997,iyer2001experimentalist} or amplitude \cite{ByMaGr14} reconstruction rely on the dynamics being nearly recurrent and break down for spatiotemporally chaotic states featuring frequent topologically nontrivial events such as the creation or annihilation of spiral cores.}

A more conventional (and convenient) approach is to use {instead} the spiral tip, which is a point on the boundary $\partial E$ of the excited region that separates the depolarization wavefront from the repolarization waveback (green circle in Fig. \ref{fig:anatomy}(b)). 
A number of different definitions of the spiral tip have been introduced in the literature.
The most popular are the ones based on the level-set intersection (LSI)~\cite{BaKnTu90} \begin{align}\label{eq:lsi}
	u_1(t,\vec{x}) = \bar{u}_1,\qquad f_1^0(\bar{u}_1,u_2)=0,
\end{align}
or zero normal velocity (ZNV) \cite{Fenton1998}
\begin{align}\label{eq:znv}
	u_1(t,\vec{x}) = \bar{u}_1,\qquad \partial_tu_1(t,\vec{x}) = 0,
\end{align}
or the curvature $\kappa$ of the level set $u_1(t,\vec{x}) = \bar{u}_1$ \cite{beaumont1998spiral}.
In particular, ZNV and LSI define the spiral tip(s) $\vec{x}_{p}(t)$ as intersection of two level sets which are much easier to compute than the curvature.
LSI can be thought of as the limiting case of ZNV where $D_{11}\to 0$, so the difference between the positions of spiral tips defined using these two methods provides a measure of the importance of electrotonic effects.

Albeit they could be simple to define, spiral tips {typically} exhibit spurious dynamical effects. 
For instance, they move (in circular trajectories) for spiral wave solutions of \eqref{eq:rde} rigidly rotating around an origin $\vec{x}'$ (n\'ee relative equilibria) which satisfy
\begin{equation}\label{eq:req}
	\partial_t{\vec{u}}(t,\vec{x}) = \omega \partial_{\theta}{\vec{u}}(t,\vec{x}),
\end{equation}
where $\partial_{\theta} = \hat{\vec{z}}\cdot(\vec{x}-\vec{x}')\times\nabla$ and $\omega=2\pi/T$ is the angular frequency.
Indeed, even if the normal velocity \eqref{eq:cn} of the spiral tip vanishes, its tangential velocity will not vanish, unless $\bar{u}_1=u_1(\vec{x}')$.
Hence, spiral tips are not ideally suited to be used as indicators of the wave dynamics.

The phase singularity, unlike the spiral wave tip, should remain stationary for a rigidly rotating spiral wave. 
This requires that the location $\vec{x}_{o}(t)$ of every phase singularity satisfies
\begin{equation}\label{eq:origins}
	 \partial_t\vec{u}(t,\vec{x}_o) = \vec{0}.
\end{equation}
Equivalently, $\vec{x}_{o}(t)$ correspond to the intersections of the level sets 
\begin{equation}\label{eq:dR}
\partial R=\{\vec{x}|\partial_tu_1(t,\vec{x}))=0\}
\end{equation}
and $\partial E$ defined according to \eqref{eq:dE}, i.e., phase singularities are points on the boundary of the excited region that separate the refractory region from the excitable region.
{Note that, for $\partial E$ defined by \eqref{eq:adu}, its intersections with $\partial R$ are independent of $\alpha$, and so is the definition of the phase singularities. This is explicit in the definition \eqref{eq:origins}.}

It is easy to see that the boundaries $\partial E$ and $\partial R$ merely correspond to different level sets {$\varphi=\alpha\pm\pi/2$ and $\varphi=\pm\pi/2$} of the phase field
\begin{equation}\label{eq:phase}
	 \varphi=\mathrm{arg}(\partial_tu_1+i\partial_tu_2),
\end{equation}
so $\vec{x}_{o}(t)$ indeed corresponds to a phase singularity of the phase field \eqref{eq:phase}.
Since they are defined locally (just like the spiral wave tips defined via LSI and ZNV), the phase singularities can be easily determined for arbitrarily complicated solutions.
More generally, $\vec{x}_o(t)$ can be interpreted as the instantaneous center of rotation for slowly drifting spiral waves, i.e., for which the rotation-averaged translation of the spiral wave core is much smaller than the typical propagation velocity $c$ of excitation waves.

The positions of spiral wave tips and phase singularities are compared in Fig.~\ref{fig:WFWBcores}(c).
For the LSI and ZNV definitions, the positions of the spiral tips are shown for $1.68 \leq \bar{u}_1 \leq 2.11$, corresponding to the voltage threshold between $\APD_{70}$ and $\APD_{90}$, respectively~\cite{Fenton2002}. 
Clearly, electrotonic effects are non-negligible for the present model, as the tip positions predicted by LSI are dramatically different from those defined by ZNV over a range of choices of $\bar{u}_{1}$.
The tip positions defined by ZNV much more closely match the phase singularities defined by \eqref{eq:origins}, and mostly differ in position in the direction normal to the wavefront (along the local gradient of $u_1$).

In conclusion of this section, we should mention that, for a typical multi-spiral solution, $u_1$ varies rather significantly across the spatial domain, while $u_2$ is restricted to a narrow range of $O(\epsilon)$ width around the value $\bar{u}_2$ that corresponds to the ``stall solution'' for a planar front connecting the two stable branches of the $u_1$-nullcline
\begin{equation}\label{eq:stall}
\int_{u_1^-}^{u_1^+}\dd u_1 \, f_1(u_1,\bar{u}_2)=0,
\end{equation}
where $f_1^\pm(u_1^\pm,\bar{u}_2)=0$ (cf. Fig. \ref{fig:anatomy}(a)). 
This is a characteristic value that corresponds to the phase singularities, as shown by Fife \cite{Fife1984}.
For the parameters used here \cite{Marcotte2016} $\bar{u}_2=0.9724$.

\subsection{Topological description}\label{sec:topo}

We can associate chirality (topological charge) $q_j=\pm 1$ with each of the {phase singularities (enumerated by $j=1,2,\dots$)}, which determines whether the spiral wave rotation is counter- or clockwise.
Chirality can be defined locally \cite{li2013chiral} as 
\begin{equation}
q_j={\rm sign}(\hat{\bf z}\cdot\nabla u_1\times\nabla u_2).
\end{equation}
For spatially discretized models it is more reliable to use a nonlocal definition of chirality.
Let us define a neighborhood of each phase singularity $\vec{x}_{o,j}$ using the window function 
\begin{equation}
	w_j(\vec{x}) = e^{-r_j/d_j},
\end{equation}
where $r_j=|\vec{x}-\vec{x}_{o,j}|$ and $d_j=\min_{k\neq j} |\vec{x}_{o,j}-\vec{x}_{o,k}|$ is the distance to the nearest distinct phase singularity.
Further, let us define the pseudo-chirality $\tilde{q}_j$ of each spiral wave as the value for which the function 
\begin{equation}\label{eq:chirality}
	J(\tilde{q}_j)=\int_{\Omega} \,\dd^{2}\vec{x} \, w_j(\vec{x})
	\left|\partial_t\vec{u}(t,\mathbf{x})-\tilde{q}_j\omega\partial_{\theta}\vec{u}(t,\vec{x}) \right|^{2},
\end{equation}
is minimized.
{The functional $J(\tilde{q}_j)$ defines a local reference frame rotating with angular velocity $\tilde{q}_{j} \omega$ around the phase singularity $\vec{x}_{o,j}$; this functional is minimized for spiral waves that are stationary in that reference frame.}
For a single rigidly rotating spiral wave, chirality is precisely $\pm 1$. 
Minimization of \eqref{eq:chirality} for complex multi-spiral states produces pseudo chirality values equal to $\pm 1$ within a few percent, such that we can safely define $q_j={\rm sign}(\tilde{q}_j)$.
{In practice, this definition proves very robust when spiral cores are sufficiently well separated, i.e., when $d_j$ exceeds the width of an isolated spiral core~\cite{Marcotte2016b}.}

Since phase singularities by definition \eqref{eq:origins} lie on the level set $\partial E$, for periodic boundary conditions, wavefronts and wavebacks can only terminate at a spiral core (or, more precisely, phase singularity).
Conversely, in multi-spiral states, each wavefront and waveback is bounded by a pair of spiral cores of opposite chirality.
In modern electrophysiology literature wavelets are identified with wavefronts \cite{Chen2000}.
Consequently, the events when a wavelet is created (destroyed) are associated with an increase (decrease) in the number of spiral cores by two.
Although the total number of spiral cores is not conserved, the total topological charge 
\begin{equation}\label{eq:topocons}
q=\sum_jq_j=0
\end{equation}
is conserved \cite{davidsen2004,Zhang2007}.

A number of topologically distinct processes which respect \eqref{eq:topocons} are possible.
Although some of these correspond to the time-reversed version of the others, the dynamics of the dissipative systems are not time-reversible and do not have to respect the symmetry between these processes. 
In fact, as we will see below, in excitable systems such as the Karma model the dominant topological processes increasing/decreasing the number of spiral cores are not related by time-reversal symmetry.

\section{Conduction block and tissue refractoriness}\label{sec:block}

As we mentioned previously, conduction block plays a major role in wave breakup, which is essential for transition to fibrillation and spiral wave chaos in general. 
The origin of conduction block can be structural, i.e., related to tissue heterogeneity \cite{courtmanche91}, but it can also be dynamical, i.e., occur in homogeneous tissue as a result of an instability.
For instance, conduction block can occur when a receding waveback is moving slower than the subsequent advancing wavefront \cite{Courtemanche:1996dl}. 
In fact, there is a variety of other dynamical mechanisms leading to conduction block \cite{Fenton2002}.

Conduction block refers to the failure of an excitation front to propagate because the tissue ahead of it is refractory and cannot be excited. 
Refractoriness is traditionally~\cite{merideth1968electrical} defined on the level of individual cardiac cells by quantifying whether a voltage perturbation applied to the quiescent state of the cell will trigger an action potential.
These definitions are not particularly useful for understanding conduction block in tissue for two reasons~\cite{Starmer1993,Starobin1994,Neu1997}:
First of all, in tissue the excitation wave is triggered by coupling between neighboring cells, rather than a voltage perturbation to an isolated cell. 
Second, in tissue, especially during AF or tachycardia, which are both characterized by very short $\DI$s, the cells never have sufficient time to return to the rest state.

\subsection{Low-curvature wavefront}

In the Karma model, conduction block can arise as a result of discordant alternans instability \cite{Echebarria2002,echebarria2007} which leads to variation in the width and duration of action potentials. 
For excitation waves with low curvature, we can determine the boundary of the refractory region by considering a one-dimensional periodic pulse train.
In the reference frame moving with velocity $c$ of the wavefront, the {voltage variable satisfies the evolution equation} 
\begin{equation}\label{eq:cmf}
	D_{11} u''_1 + c u'_1 + f_1(\vec{u}) = 0,
\end{equation}
{provided the pulse train does not change shape}, where $u'_1 = \partial_{\xi}u_1(\xi)$, $u''_1 = \partial_{\xi}^{2}u_1(\xi)$, and $\xi = x-ct$.
For sufficiently small $\DI$, the conduction velocity $c$ decreases monotonically with $\DI$ and vanishes identically at finite $\DI$~\cite{karma94}.
This means that there are no propagating solutions below this value of $\DI$. 
At the critical value of $\DI$, we have $c=0$, so the wavefront fails to propagate when $D_{11}u''_1 + f_1(\vec{u}) = 0$.
For plane waves in two dimensions, $u''_1=\nabla^2 u_1$, so combining this with the evolution equation \eqref{eq:rde} we find that the boundary of the refractory region is given by
\begin{equation}\label{eq:CB}
\partial_tu_1=D_{11}\nabla^2 u_1+f_1(\vec{u}) = 0
\end{equation}
and coincides with the boundary \eqref{eq:dR} of the repolarization region.
Similarly, the refractory region can be identified with the region of repolarization \eqref{eq:R}.
This makes intuitive sense: whatever the conditions are, the voltage increases outside the refractory region.

\begin{figure}
	\begin{center}
		\includegraphics[width=\columnwidth]{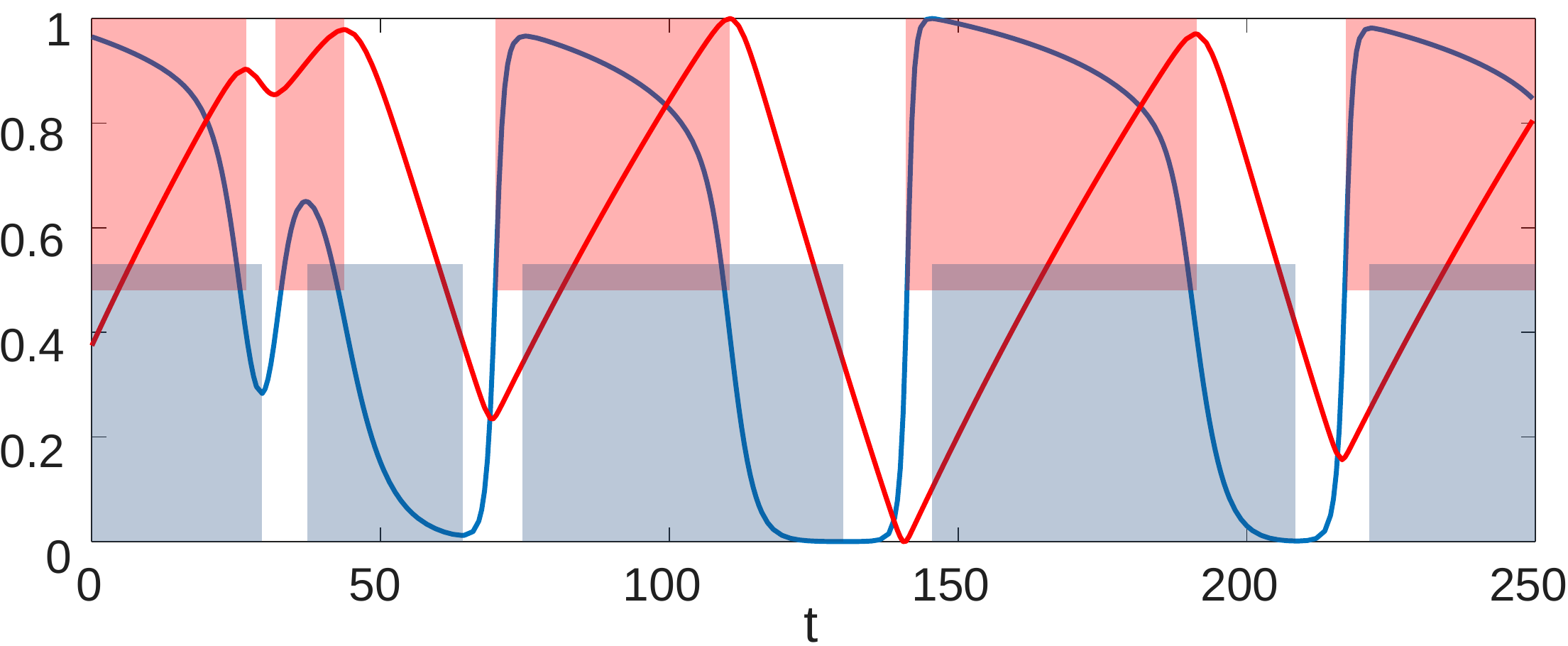}
	\end{center}
	\caption{ Time trace of variables $u_1$ (blue curve) and $u_2$ (red curve) at a fixed spatial location in a two-dimensional domain near which a conduction block occurs at $t\approx 33$. Both $u_1$ and $u_2$ were rescaled so that their range corresponds to $[0,1]$. Excited intervals are shaded red and refractory intervals are shaded blue. The first refractory interval ends at $t\approx 26.5$, just before the next excited interval begins at $t\approx 29.5$, leading to a short, small-amplitude action potential. 
	\label{fig:t-trace}}
\end{figure}

Although derived for a very special case of one-dimensional periodic pulse trains, this definition of the refractory region works well even for states that are not time-periodic and feature excitation waves with significant curvature.
This is illustrated in Fig.~\ref{fig:t-trace} which shows the time trace of of the variables $u_1(t,\vec{x}_0)$ and $u_2(t,\vec{x}_0)$ for a spatiotemporally chaotic solution similar to that shown in Fig. \ref{fig:WFWBcores}. 
The point $\vec{x}_0$ was chosen near the spatial location where conduction block occurs {(such as the center of the marked region in Fig. \ref{fig:coal} below).}
The excited and refractory intervals (temporal analogues of the excitable and refractory regions) are shown as red- and blue-shaded rectangles {in Fig. \ref{fig:t-trace}; they are bounded by the level sets $\partial R$ and $\partial E$ in Fig. \ref{fig:coal}}.

As expected, we find that, when the wavefronts are well separated from the trailing edges of the refractory intervals (e.g., at $t\approx 140$), long, large-amplitude action potentials are found.
This is in sharp contrast with the short and low-amplitude action potential that is initiated at $t\approx 29.5$, soon after the previous refractory interval ends at $t\approx 26.5$. As the location of conduction block is approached (not shown), the separation between the trailing edge of the refractory interval and the subsequent excitation wavefront vanishes along with the action potential itself. This suggests that conduction block occurs when and where the level sets $\partial R$ and $\partial E$ first touch, in agreement with Winfree's critical point hypothesis \cite{Winfree1989}. 

\subsection{High-curvature wavefront}\label{sec:curvature}

There are, however, other dynamical mechanisms that can lead to conduction block.
Consider, for instance, the opposite situation when the curvature of the wavefront is high.
For curved wavefronts the propagation speed $c$ decreases as the curvature $\kappa$ of the wavefront increases, so there is a critical value of the curvature at which the wave fails to propagate~\cite{fast1997role}.
It can be estimated in the limit $D_{22}\rightarrow 0$ using the eikonal approximation \cite{Keener:1986ag} which gives
\begin{equation}\label{eq:eikonal}
c = c_0 - D_{11}\kappa, 
\end{equation}
where $c_0$ is the velocity of a planar wavefront.
Using the value of $c_0 = \lambda/T \approx 1.44$ which corresponds to a large rigidly rotating spiral {wave~\cite{Marcotte2015}}, we find $\kappa^{-1}=r_c \approx 2.8$, where $r_c$ is the critical radius of curvature of the wavefront.

A more accurate estimate for $r_c$ can be obtained using the condition \eqref{eq:CB} for conduction block
\begin{align}\label{eq:bvp}
  D_{11}[\partial_r^2u_1 + r^{-1}\partial_ru_1 + r^{-2}\partial_{\theta}^2u_1] + f_1(\vec{u})=0
\end{align}
and the definition of the wavefront \eqref{eq:dE} rewritten via \eqref{eq:rde}
\begin{align}\label{eq:wf}
  D_{22}[\partial_r^2u_2 + r^{-1}\partial_ru_2 + r^{-2}\partial_{\theta}^2u_2] + f_2(\vec{u})=0
\end{align}
in polar coordinates $(r,\theta)$.
For small $\epsilon$, $u_2$ varies slowly in both time and space and can be considered a constant,  $u_2=\bar{u}_2$ given by \eqref{eq:stall}.
The first three terms of \eqref{eq:wf} can therefore be ignored, so \eqref{eq:wf} reduces to 
$f_2(\vec{u})=0$.
The third term $D_{11}r^{-2}\partial_{\theta}^2u_1$ in \eqref{eq:bvp} can also be neglected, since $u_1$ varies much faster in the direction normal to the wavefront $r=r_c$ than in the tangential direction. 
Since $u_2$ is constant, subject to boundary conditions $\partial_ru_1=0$ at $r=0$ and $r=\infty$, \eqref{eq:bvp} has rotationally symmetric solutions $u_1=u_1(r)$ with a stationary wavefront at the critical radius $r_c$ given by
\begin{equation}\label{eq:rc}
f_2(u_1(r_c),\bar{u}_2)=0.
\end{equation}
 
\begin{figure}[tpb]
	\begin{center}
		\includegraphics[scale=1]{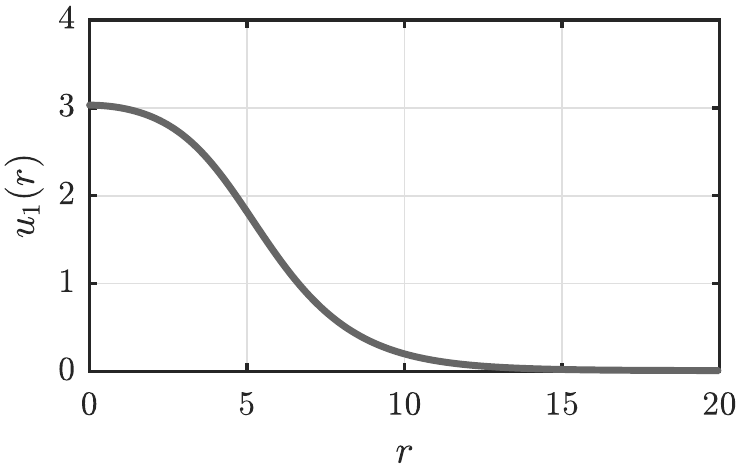}
	\end{center}
	\caption{The stationary solution $u_1(r)$ of \eqref{eq:bvp}.
	\label{fig:bumpbvp}}
\end{figure}

The stationary solution of \eqref{eq:bvp} and \eqref{eq:rc} is shown in Fig.~\ref{fig:bumpbvp}. It corresponds to the critical radius of curvature $r_c\approx 6$, which is a factor of two larger than the value obtained using the eikonal approximation \eqref{eq:eikonal}.
This solution is a two-dimensional analogue of the one-dimensional ``critical nucleus'' for excitation~\cite{Neu1997}.
Wavefronts with radius of curvature larger than $r_c$ propagate forward, while the wavefronts with radius of curvature smaller than $r_c$ retract (i.e., become wavebacks).

Before we discuss the numerical results, let us emphasize that, with the proper choice of variables, the definitions of the wavefronts and wavebacks \eqref{eq:dE}, leading and trailing edges of the refractory region \eqref{eq:dR}, and phase singularities \eqref{eq:origins} are model-independent and can be used to analyze both numerical and experimental data, provided that  measurements of two independent variables (e.g., voltage and calcium) are available.
Generalization of the topological description presented above to higher-dimensional models is discussed in the Appendix.

\section{Numerical results}\label{sec:results}

As mentioned in the introduction, during AF most spiral waves do not complete a full rotation.
Spiral wave chaos in the Karma model produces qualitatively similar dynamics: topological changes involving a change in the number of spiral wave cores occur on the same time scale as the rotation.
(For parameters considered in this study rotation period is $T\approx 51$, which corresponds to $127$~ms in dimensional units \cite{Marcotte2015}.)
The larger the spatial domain, the more frequent are the topological changes in the structure of the solution. 
However, as we discussed in Sect. \ref{sec:topo}, each topological event is essentially local and involves either birth or annihilation of a pair of spiral cores of opposite chirality. 

Of the different types of topological events, spiral wave breakup -- associated with a creation of a new pair of spiral cores -- received a lion's share of the attention due to its role in the initiation of fibrillation.
However, the number of spiral cores cannot increase forever; eventually a dynamic equilibrium is reached when the number of cores fluctuates about some average, with core creation balanced by core annihilation.
To the best of our knowledge, the process(es) responsible for core annihilation, however, have never been studied systematically.
To investigate which of the topological events dominate and what the dynamical mechanisms underlying these events are, we performed a numerical study of the Karma model \eqref{eq:rde}-\eqref{eq:karmakinetics} on a square domain of side-length $L=192$ ($5.03$~cm), {which is close to the minimal size required to support spiral wave chaos}.
Spatial derivatives were evaluated using a second-order finite-difference stencil and a fourth-order Runge-Kutta method was used for time integration~\cite{Marcotte2015}.
To avoid spurious topological transitions involving a boundary, periodic boundary conditions were used, unless noted otherwise.

\begin{figure}[tpb]
\begin{center}
	\includegraphics[width=\columnwidth]{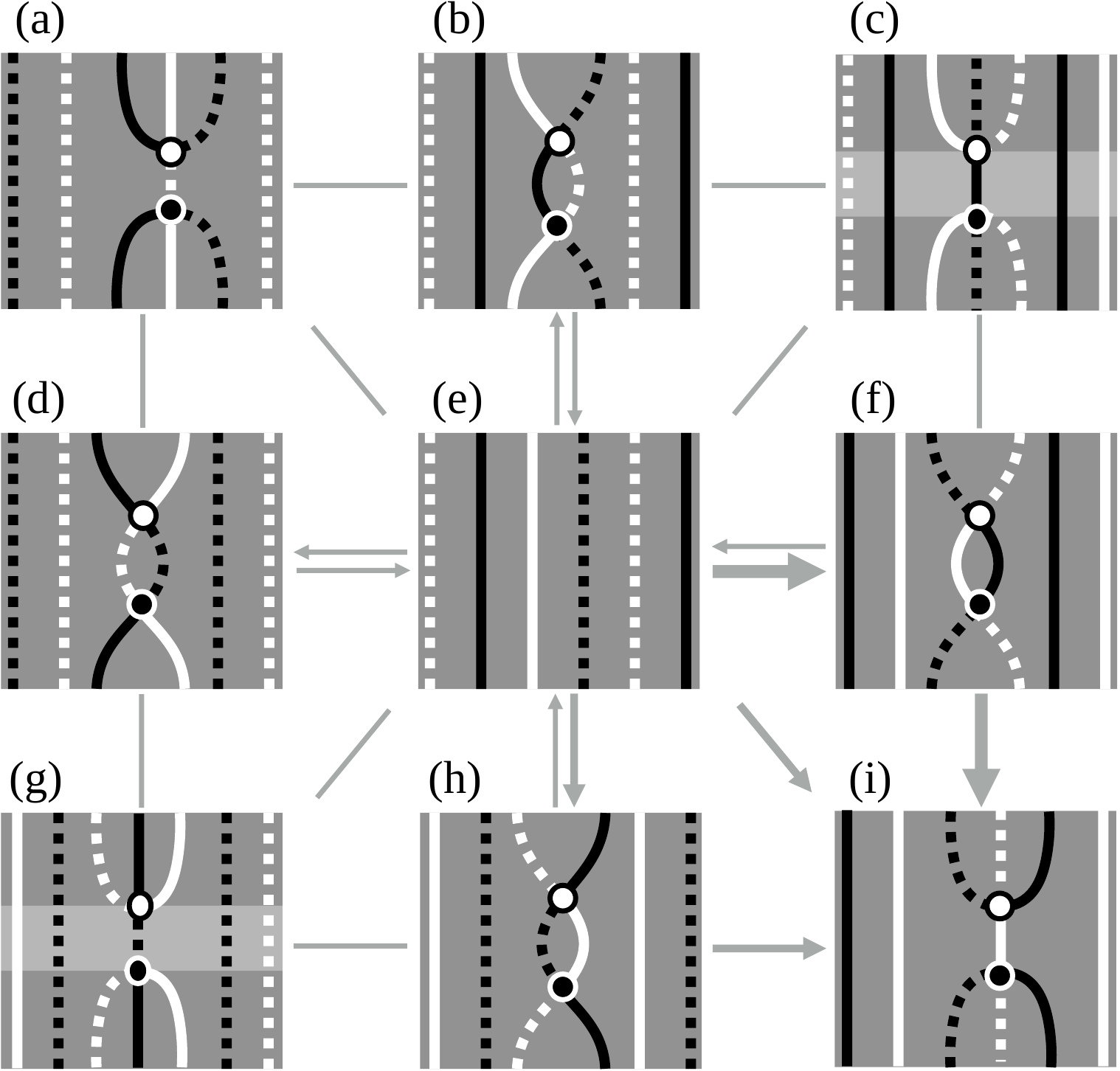}
\end{center}
\caption{Topologically distinct configurations with transitions which tend to increase the number of wavelets and cores.
The wave train moves from right to left.
Solid and dashed black lines represent, respectively, the wavefront $\partial E^+$ and waveback $\partial E^-$. 
Solid and dashed white lines represent, respectively, the leading edge $\partial R^+$ and trailing edge $\partial R^-$ of the refractory region.
The phase singularities of opposite chirality are shown as white (black) circles.
Gray lines show topologically allowed transitions that are not observed, gray arrows -- those that do occur, with thickness reflecting their frequency. {Slow transitions associated with figure-8 re-entry are not shown.}
\label{fig:master+}}
\end{figure}

Before identifying topological transitions in the numerical simulations, it is worth enumerating the topologically distinct local configurations. 
In the following it will be convenient to use the following shorthand notations: $\partial E^+$ (wavefront, $\partial_t u_1>0$), $\partial E^-$ (waveback, {$\partial_t u_1 < 0$}), $\partial R^+$ (leading edge of the refractory region, $\partial_t u_2>0$), and $\partial R^-$ (trailing edge of the refractory region, $\partial_t u_2<0$).
For a wave train in the region with no spiral cores, the boundaries of the refractory and excitable regions will follow the periodic sequence $(\dots$, $\partial R^-$, $\partial E^-$, $\partial R^+$, $\partial E^+$, $\dots)$ in the co-moving frame (cf. Fig.~\ref{fig:master+}(e)).

\subsection{Virtual pairs} 

Deformation of {(nearly planar)} waves due to instability or heterogeneous refractoriness can lead to intersection of any pair of adjacent level sets (e.g., $\partial E^+$ and $\partial R^-$) and, correspondingly, creation of a new pair of spiral cores. 
From the topological perspective, there are four distinct possibilities shown in Figs.~\ref{fig:master+}(b), \ref{fig:master+}(d), \ref{fig:master+}(f), or \ref{fig:master+}(h). 
The {corresponding configurations are all transient} and only persist for a fraction of the revolution time $T$ of a typical spiral wave {during topological transitions}. 
Each of these transient configurations can undergo a total of five distinct topological transitions, including a reverse transition back to the initial configuration with nonintersecting level sets, associated with the destruction of the two new phase singularities.
The other four persistent possibilities will be considered in subsequent sections.

The transitions that correspond to crossing of two level sets followed by the reverse transition (indicated by horizontal or vertical double gray arrows in Fig. \ref{fig:master+}) produce a ``virtual'' core pair that appears and quickly disappears, restoring the original topological structure.
The number of spiral cores and wavelets before and after these transitions remains exactly the same, so while such events do occur rather frequently they do not play a dynamically important role and can be safely ignored.
As discussed below, the topological transitions identified with the arrows in Fig. \ref{fig:master+} are all very fast; they occur on a time scale much shorter than the typical rotation period of a spiral. The much slower transitions between panels \ref{fig:master+}(a) $\to$ \ref{fig:master+}(b) $\to$ \ref{fig:master+}(c) $\to$ \ref{fig:master+}(f) $\to$ \ref{fig:master+}(i) $\to$ \ref{fig:master+}(h) $\to$ \ref{fig:master+}(g) $\to$ \ref{fig:master+}(d) $\to$ $\cdots$ associated with figure-8 re-entry for two well-separated counter-rotating spirals are not shown for clarity.
{While the topology of some level sets (either $\partial E$ of $\partial R$) changes during these transitions, neither the topological charge nor the number of phase singularities does, so these are not proper topological transitions, as defined in Section \ref{sec:topo}. In contrast, the topology of $\partial E$ and $\partial R$ may not change during the proper topological transitions.}

\begin{figure}[tpb]
\begin{center}
	\subfigure[]{\includegraphics[scale=0.9]{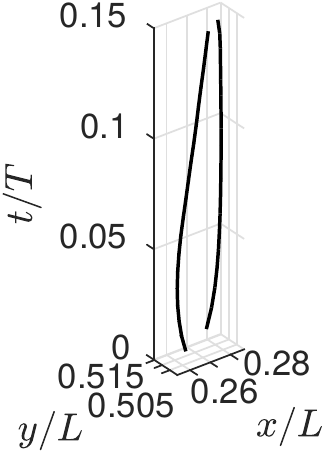}}
	\subfigure[]{\includegraphics[scale=0.85]{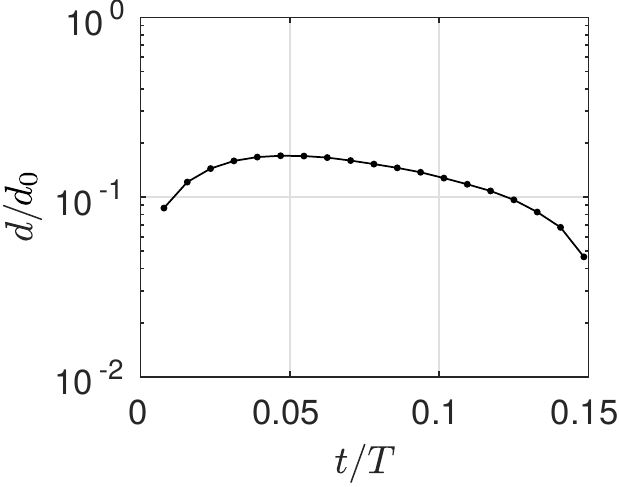}}
\end{center}
\caption{Trajectories $\vec{x}_{o}(t)$ for a virtual pair of phase singularities (a) and the distance between them as a function of time (b).
\label{fig:virtualcores}}
\end{figure}

The trajectories of the two phase singularities and the distance between them in a representative example of a virtual pair are shown in Fig.~\ref{fig:virtualcores}.
The phase singularities do not move far from their initial positions and remain close at all times. 
In fact, the distance between them never exceeds a fraction of the typical separation $d_0$ between persistent spiral cores (to be be discussed in more detail in Sect. \ref{sec:spacing}).
Since the level sets are smooth curves, their intersections that define the positions of the cores move with infinite velocities at the time instants when the cores are created and destroyed. 
As a result, their motion near those times can not be resolved using time stepping, explaining the gaps at the beginning and the end of the trajectories in Fig. \ref{fig:virtualcores}(a).

The transient configuration shown in Fig. \ref{fig:master+}(f) deserves a special mention.
It corresponds to the phenomenon of ``back-ignition'' observed in some reaction-diffusion models whereby the waveback can become a source of a new backward propagating wave under appropriate conditions. 
While topologically permissible, this configuration is {only observed as a very short transient} in the present model, reflecting the asymmetry of initial conditions imposed by the dynamics.
The relative likelihood of the four transient configurations shown in Figs.~\ref{fig:master+}(b), \ref{fig:master+}(d), \ref{fig:master+}(f), and \ref{fig:master+}(h) can in principle be computed using stability analysis of a planar wave train solution for any tissue model, but this is outside the scope of the present study.

\subsection{Wavelet/pair creation}

Next consider {the transitions} from the four intermediate configurations shown in Figs. \ref{fig:master+}(b), \ref{fig:master+}(d), \ref{fig:master+}(f), and \ref{fig:master+}(h) to {configurations} other than the initial {one} shown in Fig. \ref{fig:master+}(e). 
For each of these four transient configurations there are four distinct possibilities. 
Two possibilities are shown in Fig. \ref{fig:master+} (the other two will be discussed in the next Section): either of the two level set fragments connecting the cores can reconnect with the neighboring level set of the same type.
For instance, the configuration shown in Fig. \ref{fig:master+}(h) can transform to the configuration shown in Figs. \ref{fig:master+}(g) or \ref{fig:master+}(i). 
If the crossing and reconnection occur simultaneously, the {transition occurs directly} between the persistent configuration shown in Fig. \ref{fig:master+}(e) and one of the persistent configurations shown in Figs. \ref{fig:master+}(a), \ref{fig:master+}(c), \ref{fig:master+}(g) or \ref{fig:master+}(i) without passing through any of the intermediate {transient configurations}.
The dynamically allowed direct transitions, as determined based on the results of numerical simulations, are shown as diagonal gray arrows.

Note that the transition {between the configurations shown in Fig. \ref{fig:master+}(e) and \ref{fig:master+}(a)} corresponds to wave breakup. 
It occurs when and where the wavefront reconnects with the waveback of the same excitation wave \cite{Fife1985,Winfree1989} as a result of conduction block.
Since this topological process increases the number of disconnected excited regions, it is quite natural to find that it plays an important role in the transition from, say, normal rhythm or tachycardia (featuring a single excitation wave) to AF (featuring many separate wavelets).
While wave breakup may be prevalent during the initial stage when AF is being established, we have not found a single instance of this topological event in our numerical simulations of sustained spiral wave chaos, casting serious doubt on the premise that wave breakup plays a dynamically important role in maintaining AF in tissue or in other models.

\begin{figure}[tpb]
\begin{center}
	\subfigure[]{\includegraphics[scale=0.6]{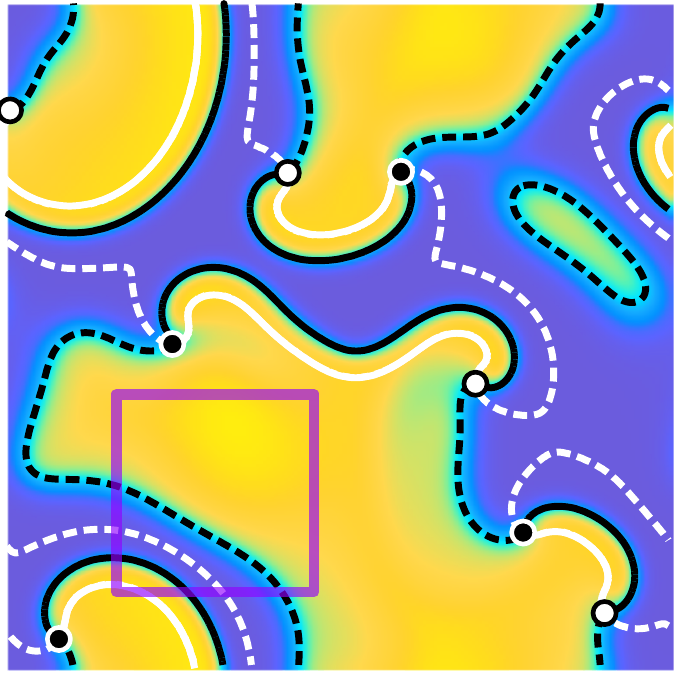}}
	\subfigure[]{\includegraphics[scale=0.6]{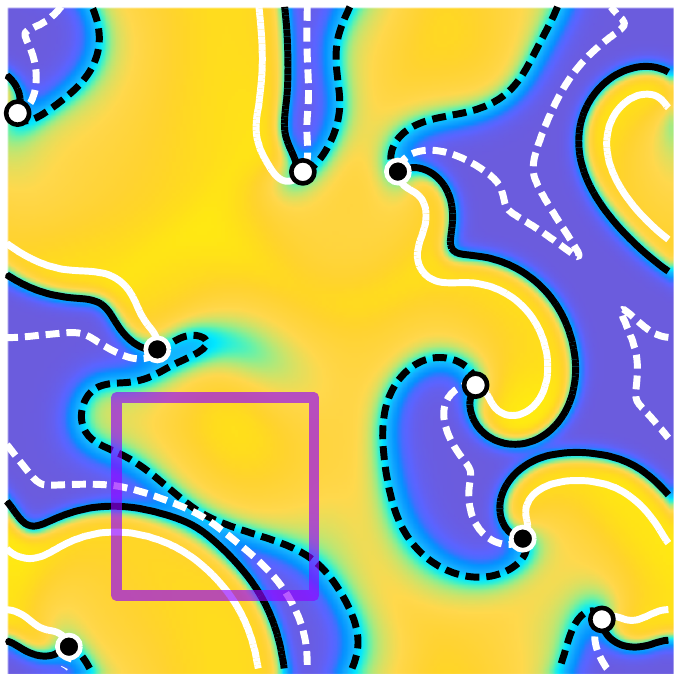}}
	\subfigure[]{\includegraphics[scale=0.6]{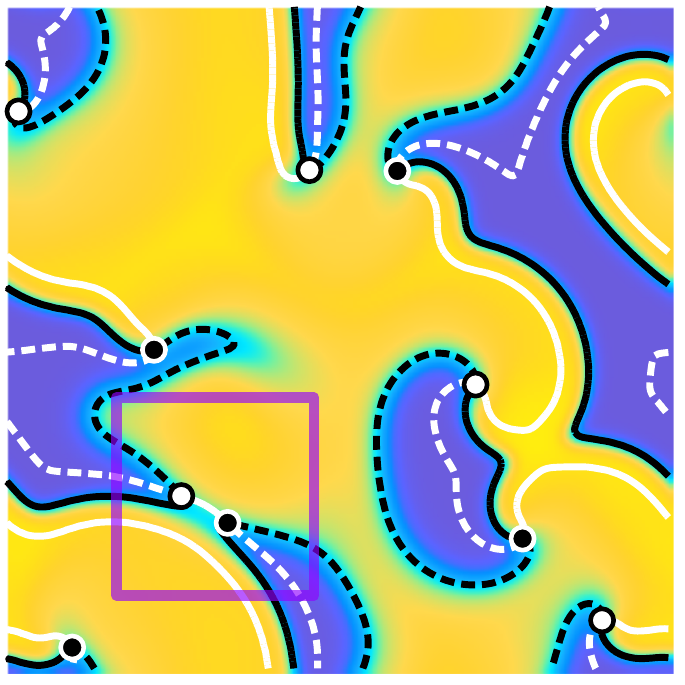}}
	\subfigure[]{\includegraphics[scale=0.6]{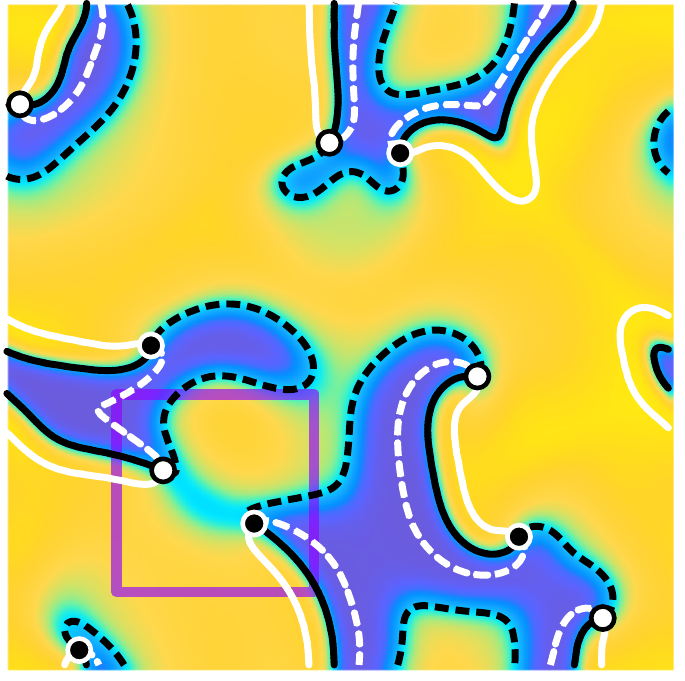}}
\end{center}
\caption{Snapshots of $u_1(t,\vec{x})$ at times $t=0.375T$, $t=0.583T$, $t=0.625T$, and $t=0.813T$ featuring wave coalescence in a chaotic multi-spiral state around $t\approx 0.60T$. The purple box indicates the region of interest. The notations are the same as in Fig. \ref{fig:master+}.
\label{fig:coal}}
\end{figure}

Our numerical simulations reveal only one topological process that leads to a lasting increase in the complexity of the pattern.
This process which we call ``wave coalescence'' corresponds to the transition from the initial configuration shown in Fig. \ref{fig:master+}(e) to the final configuration shown in Fig. \ref{fig:master+}(i), either directly or through the intermediate {transient configuration}s shown in Figs. \ref{fig:master+}(f) and \ref{fig:master+}(h). 
A representative example from the simulations is shown in Fig. \ref{fig:coal}, where a purple rectangle marks the region of interest.
Outside of this region the wavefronts are well-separated from the refractory tails, but inside the separation is markedly smaller (cf. Fig.~\ref{fig:coal}(a)).
The separation quickly decreases (cf. Fig.~\ref{fig:coal}(b)) until the level sets $\partial E^-$ and $\partial R^-$ cross and two new spiral cores with opposite chirality are created (cf. Fig.~\ref{fig:coal}(c)).
Immediately after this the two parts of the level set $\partial E^+$ reconnect, bringing the configuration to the topological state shown in Fig. \ref{fig:master+}(i).
The cores separate (cf. Fig. ~\ref{fig:coal}(c)) and the excited regions of two subsequent waves coalesce in the gap flanked by these two cores. 

\begin{figure}[tpb]
\begin{center}
	\subfigure[]{\includegraphics[scale=0.9]{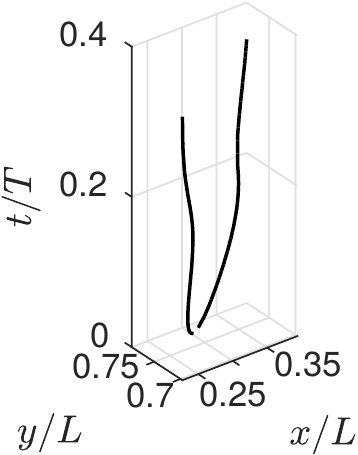}}
	\subfigure[]{\includegraphics[scale=0.85]{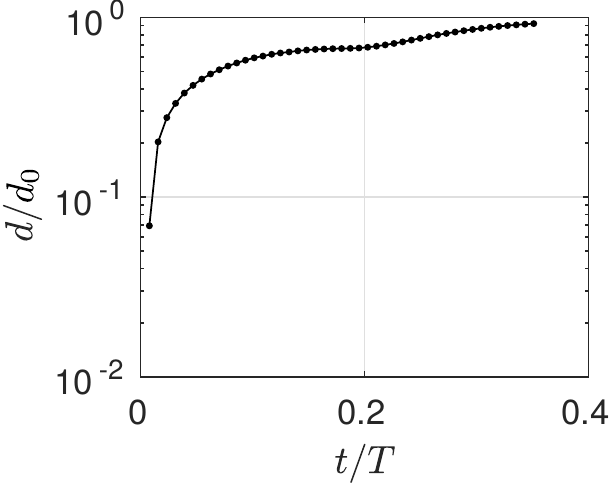}}
\end{center}
\caption{Trajectories $\vec{x}_{o}(t)$ of a pair of phase singularities post-coalescence (a) and the distance between them as a function of time (b).
\label{fig:coalcores}}
\end{figure}

Due to the high curvature of $\partial E$ the two new cores are quickly pulled apart, and two new counter-rotating spiral waves emerge, ``locking in'' the resulting topological configuration. 
This is illustrated by Fig.~\ref{fig:coalcores} which shows the trajectories of the cores and the distance between them. 
It is worth noting that, before the spiral waves complete even half a revolution, the separation between the cores approaches the typical equilibrium distance~\cite{Marcotte2016} $d_0$.

Our numerical simulations did not produce any examples of topological transitions to the configurations shown in Figs. \ref{fig:master+}(c) or \ref{fig:master+}(g). 
In the horizontal band bounded by the cores {(indicated by lighter-shade gray), the corresponding} states are characterized by the voltage variable that is changing slowly in space{, since the distance $d$ between the minimum of $u_1$ (dashed white line $\partial R^-$) and the maximum of $u_1$ (solid white line $\partial R^+$) is extremely large.
Hence, the term $D_{11}\nabla^2u_1\propto d^{-2}$} in \eqref{eq:rde} is negligible.
Since $D_{22}$ is small, we can ignore the diffusive terms $D_{22}\nabla^2u_2$ as well and consider all cells in this region to be spatially decoupled, such that their dynamics is described well by \eqref{eq:ode} and the phase diagram shown in Fig. \ref{fig:anatomy}(a). 
Consider the part of the band where $u_1$ is slowly and monotonically increasing in time (to the left of $\partial R^+$ in Fig. \ref{fig:master+}(c)) or decreasing in time (to the left of $\partial R^-$ in Fig. \ref{fig:master+}(g)).
The cells in this region should be in the state that lies close to either one of the stable $u_1$-nullclines ($f_1^-=0$ in the former case and $f_1^+=0$ in the latter case).
According to Fig. \ref{fig:anatomy}(a), this entire region should lie either to the left or to the right of the $u_2$-nullcline, so $\partial_t u_2$ should be sign-definite, while in both Fig. \ref{fig:master+}(c) and \ref{fig:master+}(g) the sign of $\partial_t u_2$ changes {(when the level set $\partial E$ connecting the two phase singularities is crossed)}.
Hence, while these configurations are not forbidden on topological grounds, they are forbidden dynamically.

Furthermore, we have not observed transitions from the persistent configurations shown in Figs. \ref{fig:master+}(a), \ref{fig:master+}(c), \ref{fig:master+}(g), and \ref{fig:master+}(i) to either the transient configurations shown in Figs. \ref{fig:master+}(b), \ref{fig:master+}(d), \ref{fig:master+}(f), and \ref{fig:master+}(h) or the persistent configuration shown in Fig. \ref{fig:master+}(e). 
While these transitions are allowed topologically, they {appear to be} forbidden dynamically. 
The only dynamically allowed (direct or indirect) transition irreversibly transforms the configuration with no spiral cores (Fig. \ref{fig:master+}(e)) to the configuration (Fig. \ref{fig:master+}(i)) with two spiral cores, increasing the total number of cores by two and the number of wavelets by one.

\subsection{Wavelet/pair destruction}

\begin{figure}[tpb]
\begin{center}
	\includegraphics[width=\columnwidth]{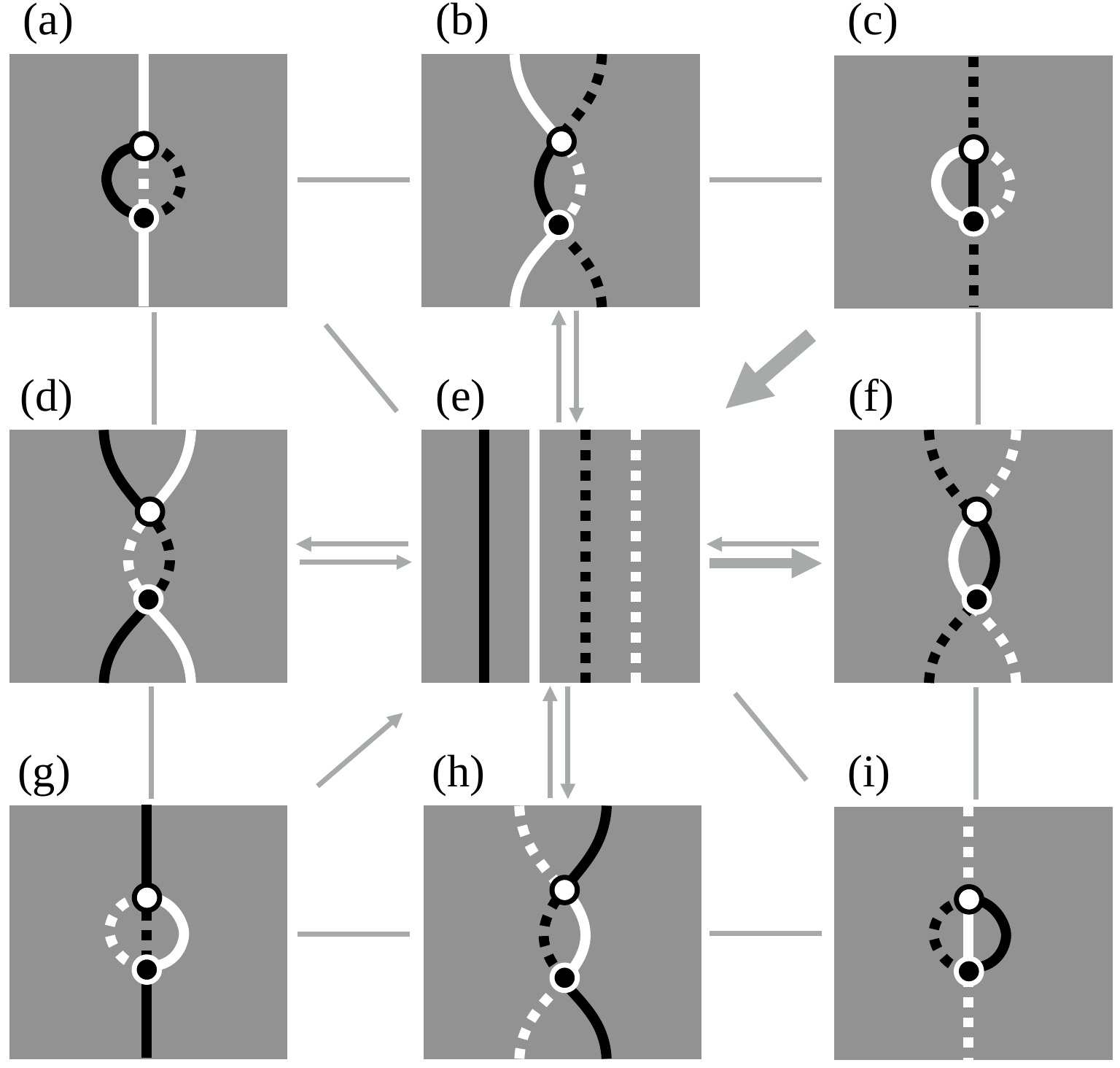}
\end{center}
\caption{Topologically distinct configurations with transitions that tend to decrease the number of wavelets and cores.
Notations are the same as in Fig. \ref{fig:master+}. {Slow transitions associated with figure-8 re-entry are not shown.}
\label{fig:master-}}
\end{figure}

Finally, let us consider the topological transitions from the four intermediate configurations that have not been considered in the previous sections.
We have redrawn these four configurations in Fig. \ref{fig:master-} in the same locations as in Fig. \ref{fig:master+}, dropping all non-essential level sets.
For each of these intermediate configurations there are two possibilities which involve reconnection between the two extended branches of a level set that terminate at the cores.
For instance, the configuration shown in Fig. \ref{fig:master-}(h) can transform to the configurations shown in Figs. \ref{fig:master-}(g) or \ref{fig:master-}(i).
None of these transitions (shown as horizontal or vertical gray lines), in either the forward or the reverse direction, have been observed in numerical simulations, however.

As we discussed previously, if the crossing and reconnection of the level sets occur simultaneously, the configuration transitions directly between the persistent configuration with no level set intersections (Fig. \ref{fig:master-}(e)) and one of the persistent configurations with a pair of intersections shown in Figs. \ref{fig:master-}(a), \ref{fig:master-}(c), \ref{fig:master-}(g), and \ref{fig:master-}(i) without passing through any of the intermediate configurations.
The dynamically allowed direct transitions observed in the simulations are shown as diagonal gray arrows.
Note that again there is no time-reversal symmetry: only the transitions that destroy the existing core pairs are dynamically allowed.
Therefore the observed direct transitions shown in Fig. \ref{fig:master-} reduce the net number of spiral cores and wavelets balancing the increase due to wave coalescence.

The configurations shown in Figs. \ref{fig:master-}(a), \ref{fig:master-}(c), \ref{fig:master-}(g), and \ref{fig:master-}(i) all describe a pair of counter-rotating spiral waves.
In particular, the configurations in Figs. \ref{fig:master-}(c) and \ref{fig:master-}(g) correspond to multi-spiral {states} (wavefronts and/or wavebacks connect spiral cores inside and outside the region shown) and hence are quite typical. 
On the other hand, the configurations in Figs. \ref{fig:master-}(a) and \ref{fig:master-}(i) correspond to configurations with a single pair of spirals (wavefronts and wavebacks connect the phase singularities inside the region shown) and are never observed during sustained spiral wave chaos.
Consequently, only transitions from the configurations in Figs. \ref{fig:master-}(c) and \ref{fig:master-}(g) are found in the simulations, with the vast majority of transitions involving the former configuration.

To understand why and when this transition happens, consider the interaction between a pair of isolated counter-rotating spiral waves separated by distance $d$.
(The interaction is short-range, so the presence of other, remote, spiral wave cores does not change the outcome.)
Using the approximate mirror symmetry of the configuration, the dynamics can be understood by considering a single spiral interacting with a planar no-flux boundary at a distance $\zeta=d/2$. 
As we showed previously \cite{Marcotte2015,Marcotte2016}, at large separations the spiral cores can be considered essentially non-interacting, while at smaller separations the equilibrium distance $d$ becomes quantized, with the smallest stable separation \cite{Marcotte2016} equal to $d_0=2\zeta_0\approx 40$ (10.4 mm in dimensional units) for the values of the parameters considered in this study.
For separations below some critical distance $d_c<d_0$, the cores attract each other, eventually colliding and destroying both spiral waves.
As the cores approach each other, the wavefront confined between them collapses, so we will refer to this process as ``wavelet collapse'' or ``wave collapse''.

\begin{figure}[tpb]
\begin{center}
\subfigure[]{\includegraphics[scale=0.6]{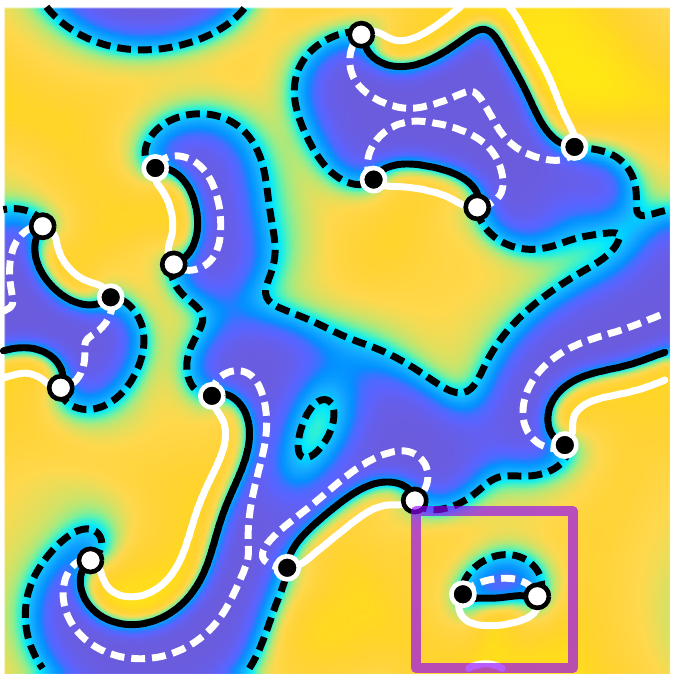}}
\subfigure[]{\includegraphics[scale=0.6]{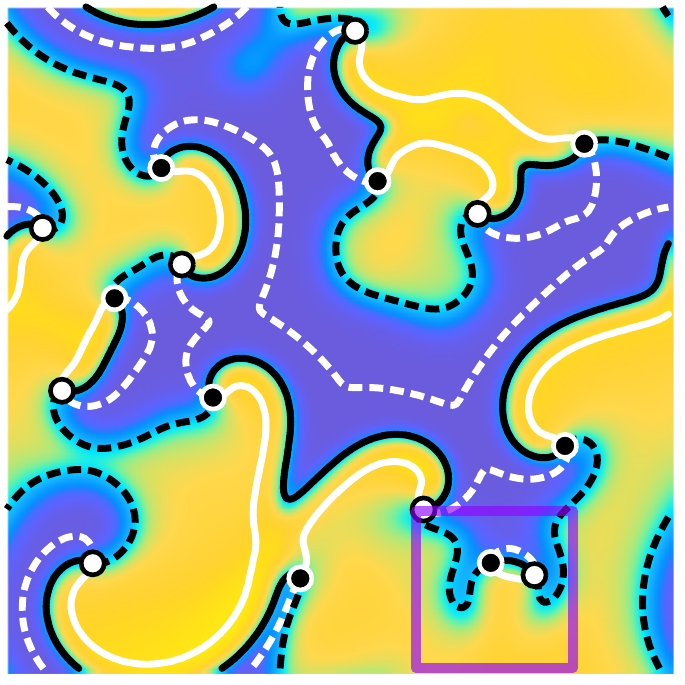}}
\subfigure[]{\includegraphics[scale=0.6]{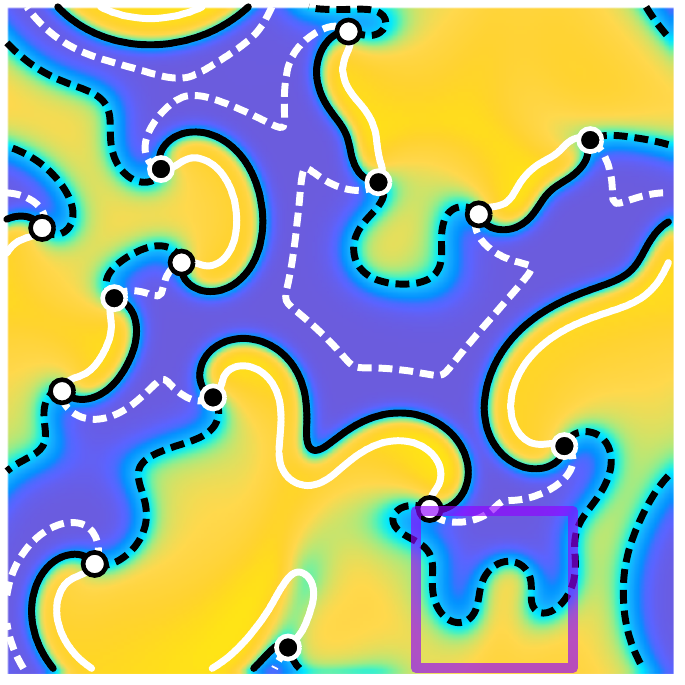}}
\subfigure[]{\includegraphics[scale=0.6]{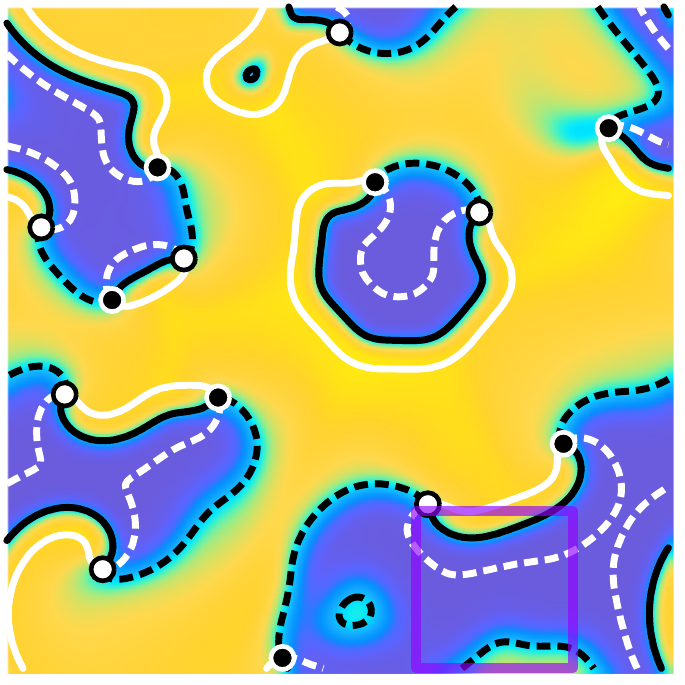}}
\end{center}
\caption{Snapshots of $u_1(t,\vec{x})$ at times $t=0.206T$, $t=0.427T$, $t=0.500T$, and $t=0.720T$ featuring wave collapse due to conduction block in a chaotic multi-spiral state around $t\approx 0.5T$. The purple box indicates the region of interest. The notations are the same as in Fig. \ref{fig:master+}.
\label{fig:merger}}
\end{figure}

The details of wave collapse depend on the relation between the initial phase of the spiral waves  and separation between their cores.
A very typical example {of wave collapse} is shown in Fig.~\ref{fig:merger}.
In this particular example we find that the curvature of the wavefront becomes quite large before collapse.
The curvature at which this happens can be related to the mechanism of conduction block discussed in Sect. \ref{sec:curvature}.
Since the cores are moving relatively slowly {prior to} wave collapse (cf. Fig.~\ref{fig:merger}(a)), as the wavefront propagates its curvature gradually increases (cf. Fig.~\ref{fig:merger}(b)).
The largest value of the curvature is related to the distance between the cores, $\kappa^{-1}\approx d/2$.
Once the curvature becomes comparable to the inverse of the critical radius $r_c\approx 6$, the wave stops propagating, the cores slide towards each other along the wavefront and annihilate (cf. Fig.~\ref{fig:merger}(c)), the wavebacks merge, and the wave starts to retract (cf. Fig.~\ref{fig:merger}(d)).

\begin{figure}[tpb]
\begin{center}
	\subfigure[]{\includegraphics[scale=0.9]{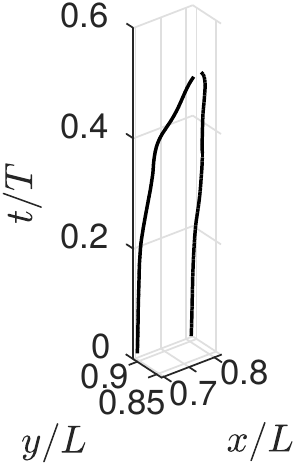}}
	\subfigure[]{\includegraphics[scale=0.85]{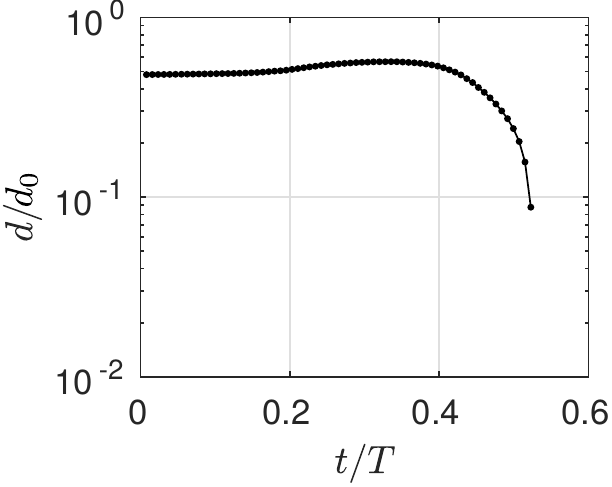}}
\end{center}
\caption{Trajectories $\vec{x}_{o}(t)$ for a pair of phase singularities during wave collapse (a) and the distance between them as a function of time (b).
\label{fig:mergercores}}
\end{figure}

This picture predicts that the minimal distance at which the spiral cores with opposite chirality can persist without annihilating with each other is given by $d_c=2r_c=12$ (3.1 mm in dimensional units). 
This value is in good agreement with the critical isthmus width (2.5 mm) found for conduction block in isolated sheets of
ventricular epicardial muscle with an expanding geometry~\cite{fast1997role}.
Our numerical simulations show that the minimal distance is $d_c=16$ (4.2 mm), also close to the predicted value.
The core trajectories and the distance between them in the example from Fig. \ref{fig:merger} are shown in Fig. \ref{fig:mergercores}.
The initial distance in this case was $d=18$ (4.7 mm), illustrating that, under appropriate conditions, wave collapse can also occur for cores separations somewhat larger than $d_c$ (but still less than $d_0$). 

The transition between the configurations shown in Figs. \ref{fig:master-}(g) and \ref{fig:master-}(e) corresponds to merger between two wavefronts that were originally separated by a waveback.
Hence, we shall refer to this topological transition as a ``wave merger'' event.
Wave mergers, however, are extremely rare, so a reduction in the total number of cores and wavelets is due almost entirely to wave collapse events.
This is similar to the dynamical asymmetry between the wave breakup and wave coalescence events.
Therefore, dynamical equilibrium in sustained spiral wave chaos can be understood, at least in the Karma model, as a balance between wave coalescence and wave collapse.

\section{Dynamical Equilibrium}\label{sec:spacing}

Although the topological description itself is not quantitative, it helps identify the key dynamical mechanisms, such as wave coalescence and wave collapse, responsible for maintaining AF.
This should, in turn, enable a quantitative description of the dynamics in general and dynamical equilibrium in particular and give the answers to open questions that have been debated for a long time.
For instance, it is presently not well understood either what the minimal size of tissue is that can sustain AF or what the minimal number of wavelets is in sustained AF. 
The leading-circle concept \cite{Allessie1977} suggests that the number of wavelets that the atria can accommodate should be related to the wavelength.
Moe's computer model \cite{Moe1964} predicted that between 23 and 40 wavelets are necessary for the maintenance of AF, while Allessie \cite{Allessie1985} places the minimal number of wavelets between four and six.

These hypotheses can be easily tested in the context of the Karma model.
Let us start by determining whether the wavelength ($\lambda=78$ for the values of parameters considered here) is a relevant length scale.
The size (diameter) of a reentry circle with the perimeter equal to the wavelength is $d=\lambda/\pi\approx25$ which is larger that the minimal separation $d_c$ between persistent spiral wave cores, but considerably smaller than the minimal stable separation $d_0$ between the cores.

\begin{figure}[tpb]
	\begin{center}
	\includegraphics[width=\columnwidth]{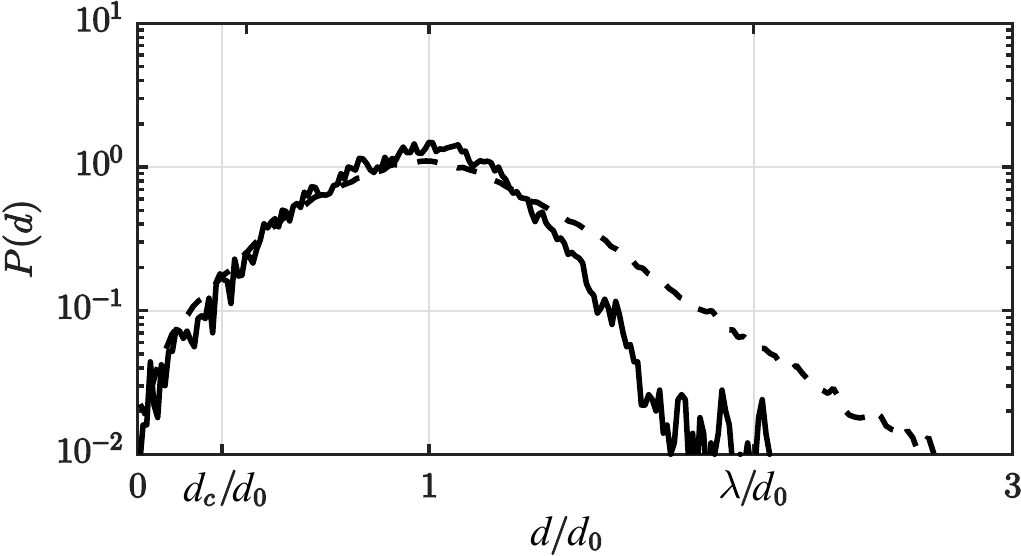}
	\end{center}
	\caption{Probability distribution of pairwise distances $P(d)$ in sustained spiral wave chaos on a square domain with side $L=192$ (5.03 cm) and no-flux boundary conditions (dashed) or periodic boundary conditions (solid).
	\label{fig:core_stats}}
\end{figure}

To show that $d_0$ is the relevant length scale, we computed the probability density function $P(d)$ for core-core separation on a square domain of side $L=192$ (this is the smallest domain with no-flux boundary conditions that supports sustained spiral wave chaos).
For each time $t$ and each core $j$ we computed the distance $d_j$ to the nearest core (cf. Sect. \ref{sec:topo}), then averaged over $j$ and $t$.
The resulting distribution, for both no-flux and periodic boundary conditions, is shown in Fig. \ref{fig:core_stats}.  
In both cases we find that the distribution $P(d)$ is rather narrow, with the maximum achieved at $d=d_0$.

\begin{figure}[tpb]
	\begin{center}
	\includegraphics[width=\columnwidth]{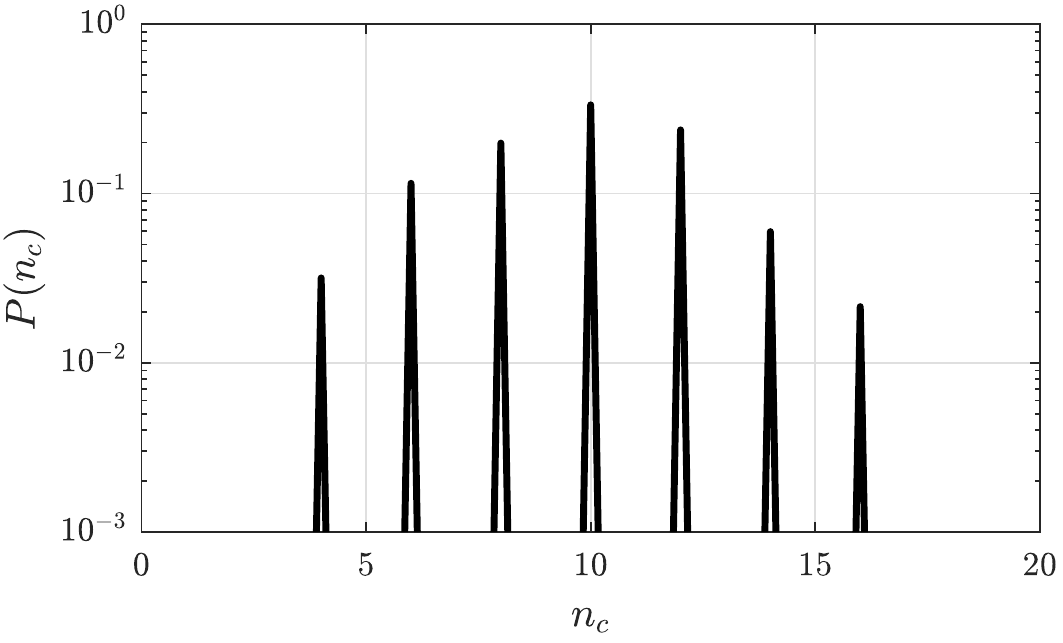}
	\end{center}
	\caption{Probability distribution of the number of cores $P(n_c)$ during sustained spiral wave chaos on a square domain with side $L=192$ (5.03 cm) and periodic boundary conditions.
	Note that odd-valued numbers of cores are forbidden by the boundary conditions.
	\label{fig:Pq2}}
\end{figure}

The effect of the boundary conditions on the shape of the distribution is somewhat subtle: on a bi-periodic domain, the probability of large core separations ($d=O(L)$) is decreased compared with the same size domain with no-flux boundary conditions.
Effectively, as there must always be a chirally-matched pair on the periodic domain, the furthest these cores may be is $d_{\rm max} = L/\sqrt{2}$, as opposed to an isolated spiral matched with it's mirror image across the no-flux boundary, which corresponds to maximal distance $d_{\rm max} = \sqrt{2}L$.
Thus, on a periodic domain, the maximal accessible distance is precisely $1/2$ the maximal distance available on a no-flux domain of the same size.

The upper bound for the number of spiral cores can be estimated as the ratio of the total area of the domain (i.e., $L^2$) to the area of the smallest tiles \cite{ByMaGr14} supporting one persistent spiral wave (i.e., $d_0^2$), that is $\bar{n}_c<L^2/d_0^2=23$ (in fact, we should have $\bar{n}_c\le 22$ since the net topological charge is zero).
In reality the tiles tend not to be squarish and have a larger area on average, giving a lower average number of spiral cores, $\bar{n}_c=10$, as the probability distribution function $P(n_c)$ illustrates (cf. Fig. \ref{fig:Pq2}).
The number $n_w$ of separate wavelets is exactly half of the number $n_c$ of cores (on a domain with periodic boundary conditions), so on average $\bar{n}_w=\bar{n}_c/2\approx 5$, in perfect agreement with the results of Allessie \cite{Allessie1985}.
The number of cores exhibits considerable fluctuation (between 4 and 16), correspondingly the number of wavelets varies between 2 and 8.
The likelihood of these extreme values is, however, rather small (an order of magnitude smaller than that corresponding to the average value).

The observation that the minimal number of wavelets is just two is a sign that the dynamics are on the border of spontaneous collapse of spiral wave chaos
{(recall that our domain is just large enough to sustain this regime).
We should have $P(0)=0$, because once all the spiral cores disappear, so does the mechanism of reentry (at least in our homogeneous model), resulting in a transition to the rest state or, in the presence of pacing, normal rhythm.
The smallest number of spiral cores required for reentry (in a domain with periodic boundary conditions) is two, so one could, in principle, expect $P(2)$ to be nonzero.
However, as our results show, the mechanism that sustains spiral wave chaos is wave coalescence, which requires at least two wavelets, and therefore at least four spiral cores, to be present.}

\section{Discussion}\label{sec:summary}

{This paper presents a general topological approach for studying spiral wave chaos in two-dimensional excitable media. 
It is illustrated using the Karma model which, in a certain parameter regime, produces dynamics that are remarkably similar to those observed during atrial fibrillation. 
Therefore, our results could shed new light on this important and complicated phenomenon.} 

The confusing and often contradictory results regarding the dynamical origins of AF reported in experimental and numerical studies are to some extent due to the complexity of the patterns of excitation. 
The descriptive language and intuition developed primarily in the context of simple structures -- plane or spiral waves -- often fail us when applied to states that are topologically complicated and nonstationary. 
To give a few examples, the mental picture of a spatially localized excitation wave, or wavelet, that is bounded by a wavefront and a waveback falls apart when applied to complex multi-spiral patterns since the boundary of one excited region is often composed of multiple wavefronts and wavebacks, as Figs. \ref{fig:coal} and \ref{fig:merger} illustrate. 
As a result, the number of excited regions almost never corresponds to the number of wavelets. 
Neither is the notion of a spiral wave immediately useful for describing such complicated patterns, which only resemble spiral waves in small neighborhoods of spiral cores. 
{Similarly, a reduction of complicated field configurations to the number and positions of phase singularities is also problematic, both because they appear, move, and disappear for sustained spiral chaos and because identifying them using existing approaches, such as phase mapping \cite{davidenko1992,Pertsov1993}, is notoriously unreliable when the data is noisy.}

This paper aims to rectify some of these difficulties by introducing a topological description that can rigorously and easily identify the dynamically important elements of the excitation patterns -- wavefronts, wavebacks, phase singularities, etc. -- without modeling assumptions and in a manner that can be implemented in both simulations and experiments. 
{By defining the phase singularities as intersections of level sets of an appropriately defined phase field, this topological description directly connects the dynamics of excitation waves and phase singularities; it} can be used not only to quantify and classify the excitation patterns, but also to identify the dynamical mechanisms that lead to qualitative changes in the pattern. 
In particular, we show that the qualitative changes can be conveniently described and classified based on the dynamics of spiral cores which are created or destroyed in pairs, leading to an increase or decrease in the number of wavelets, with a one-to-one correspondence between the number of cores and wavelets. 

The topological description also allowed us to identify the dominant dynamical mechanisms responsible for maintaining AF in a model of atrial tissue.
In particular, it allowed us to make a major discovery with implications that, in all likelihood, go far beyond the simple model considered here.
We found that wave breakup due to conduction block that is widely believed to be the key mechanism responsible for maintaining AF plays no role whatsoever in sustaining this regime.
While wave breakup does play a key role in the transition to AF, it is a dynamically and topologically distinct event -- wave coalescence -- that is responsible for maintaining AF. 
Wave coalescence which leads to the increase in the number of spiral cores and wavelets is balanced by wave collapse which decreases the number of spiral cores and wavelets.
It is this delicate balance that is responsible for maintaining the complexity of the pattern and of the dynamics and it is this balance that controls whether AF persists or terminates.
 
Past studies of the dynamical origins and control of AF tended to focus solely on the mechanism(s) that lead to an increase in the complexity of the pattern.
Indeed suppressing the processes that generate new spiral cores and new wavelets is one way to terminate or prevent AF.
However, enhancing the processes that destroy the spiral cores and wavelets could be just as effective.
Therefore, both wave coalescence and wave collapse are attractive targets for electrical, surgical, and pharmacological approaches to the treatment of AF.
While this study has not focused on the interaction of excitation waves with no-flux boundaries, the methods and approaches presented here are applicable to this situation as well.
Hence topological analysis could be quite helpful for improving treatment of chronic AF using  surgical procedures such as ablation that effectively introduce additional boundaries.

In conclusion, we should point out that our results raise new questions regarding the role of conduction block in maintaining AF. 
While conduction block undoubtedly plays a crucial role in wave collapse, it is not at all clear that it is relevant in wave coalescence.
Therefore, quite paradoxically, we find that conduction block plays a more important role in decreasing the complexity of the excitation pattern than in increasing its complexity.
Further studies are needed in order to fully understand the dynamical mechanisms behind wave coalescence, wave collapse, and possibly other topologically allowed events important in maintaining AF using more detailed and physiologically accurate models of atrial tissue.

\begin{acknowledgments}

This material is based upon work supported by the National Science Foundation under Grant No. CMMI-1028133.  
The Tesla K20 GPUs used for this research were donated by the ``NVIDIA Corporation'' through the academic hardware donation program.
CDM gratefully acknowledges the financial support of the EPSRC via grant EP/N014391/1 (UK).

\end{acknowledgments}

\appendix
\section{Application to more realistic action potentials}

The topological approach introduced here is sufficiently general to be extended to much more complicated electrophysiological models.
For example, the definition of the leading and trailing edges of the refractory region \eqref{eq:dR} relies solely on the voltage variable, while the definition \eqref{eq:adu} of the wavefronts and wavebacks can be trivially generalized to 
an $m$-variable reaction-diffusion model by choosing the ``weighting'' vector $\vec{a} = [a_1,\dots,a_m]$ to properly represent the physiological role of different variables in triggering the depolarization front.
{The phase singularities can then be defined again as the intersection of the boundaries $\partial E$ and $\partial R$, although this does not guarantee that \eqref{eq:origins} will be satisfied for $m>2$.}

\begin{figure}
	\begin{center}
	\includegraphics[scale=1.0]{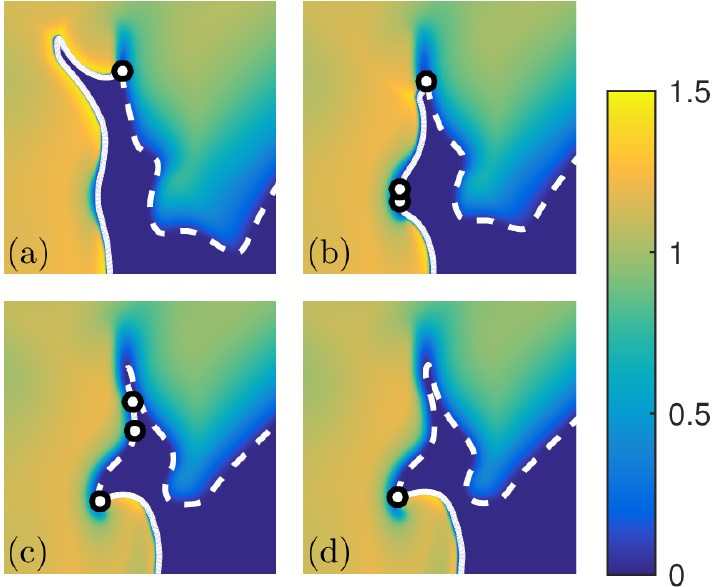}
	\end{center}
	\caption{Snapshots of the non-dimensional voltage variable $u_1(t,\vec{x})$ illustrating a virtual pair creation/destruction event in the Bueno-Orovio model~\cite{Bueno2008}. A small region of a much larger computational domain is shown with the wavefronts (solid white lines), wavebacks (dashed white lines), and phase singularities (circles) superimposed on the voltage field.
	\label{fig:bochfe}}
\end{figure}

For illustration, we used such a generalization to identify the wavefronts, wavebacks, and phase singularities in the four-variable minimal model of Bueno-Orovio {\it et al.}~\cite{Bueno2008}, considering only the contribution from a single slow variable, $\vec{a} = [0,0,1,0]$, as a simple approximation. 
A virtual pair event shown in Fig.~\ref{fig:bochfe} provides an illustration that these definitions are equally successful in describing topological changes in a substantially more complex and detailed model, which is capable of producing quantitatively accurate description of excitation patterns in various types of cardiac tissues, given appropriate choices of parameters (in the present example we used the epicardial parameter set). 

The only complication arises when the diffusion coefficients for the slow variable(s) vanish identically, since this can lead to subtle artifacts when discontinuities of the kinetics, e.g., in the switching between on and off states, combine with the high spatial gradients near the wavefront. 
However, this issue is not a fault of the method, but rather a consequence of the unphysical nature of the simplified ionic kinetics, and can be easily rectified by using an appropriately smoothed version of the model kinetics (as we did in Karma model).
Furthermore, even without smoothing, one can simply define the wavefront and waveback as the boundary of the excited region $E$ defined using the indicator function $g({\bf u})=\vec{a}\cdot\partial_t\vec{u}$.

\section*{References}
\bibliography{../bibtex/cardiac}

\end{document}